# Aging-driven in situ polymerization of FEC additive boosts the calendar-life of silicon anodes via surface passivation enhancement

*Sattajit Barua,[1] Rownak J. Mou,[1] and Koffi P. C. Yao[1*]*

[1]Department of Mechanical Engineering, University of Delaware, Newark, Delaware 19716, United States of America

KEYWORDS: Silicon anode, Calendar aging, Solid Electrolyte Interphase (SEI), Polycarbonate, Distribution of Relaxation time (DRT), Symmetric cell

**Corresponding author**: Koffi P.C. Yao, claver@udel.edu

TABLE OF CONTENT

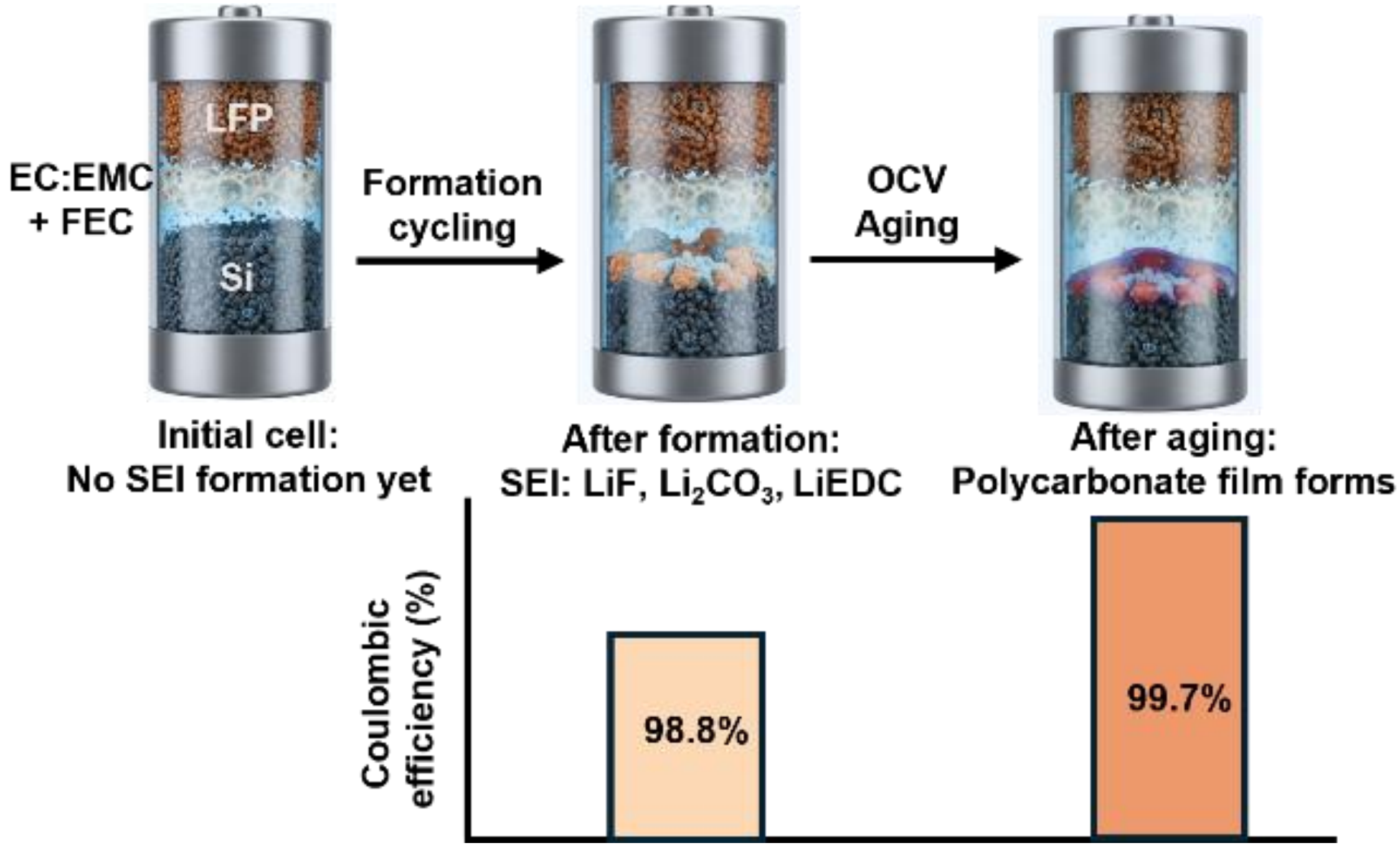

ABSTRACT

The role of additives such as FEC in extending the calendar life of silicon anodes beyond the cycling benefits is still not fully understood. Herein, the calendar life of high-loading Si (80 wt%) using baseline 1.2 M $LiPF_6$ in EC: EMC electrolyte versus adding 10 wt% FEC is investigated over months. Over 8 days of aging, FEC leads to a 13-fold reduction in irreversible capacity loss in Si||$LiFePO_4$ full cells. Cells without FEC are projected to fall below 80% of their initial capacity within ~22 days versus ~279 days with FEC. Symmetric Si||Si cells from harvested electrodes show greater increase in interphase resistance without FEC, whereby an increase of 10.81 Ω is measured for 0 wt% FEC vs. only 3.37 Ω for 10 wt% FEC over 2 months. Power law modeling of this long-term interphase resistance finds mixed transport–reaction growth behavior in FEC-free cells, suggesting significant dissolution, whereas cells with 10 wt% FEC added display a diffusion-controlled impedance growth behavior, suggesting a robust surface passivation film. Post-mortem FTIR and XPS confirm polycarbonate enrichment of the SEI, which was discovered to predominantly emerge from FEC self-polymerization during the idle aging. When the Si electrodes aged with and without FEC are harvested and reassembled into full cells with the same electrolytes used at aging, the first-cycle coulombic efficiency is 71% for 0 wt% FEC versus 97% for 10 wt% FEC. Subsequent cycling maintains over 99.7% CE with 10 wt% FEC, surpassing the pre-aging CE of 98.8%. This elevated CE indicates better passivation provided by the polymer fragments formed during aging compared to electrochemically formed SEI where no strong polymer FTIR signal is found. The self-polymerization during idle aging with additives such as FEC is therefore an opportune in situ mechanism to further engineer in extending the life of Si-based batteries.

## 1. INTRODUCTION

Silicon (Si) is a promising anode material for next-generation Lithium-ion (Li-ion) batteries due to its high theoretical specific capacity of 3579 mAh·g$^{-1}$, low reversible potential versus Li/Li$^{+}$, and crustal abundance.[1,2] However, for industrial viability, Si anodes must achieve 1000+ charge/discharge cycles and 10 years calendar life (limited capacity loss while not in use). The mechanical and chemical stability of Si electrodes is challenged by the ~350% volume expansion and contraction of Si particles during cycling, which causes stress and fatigue failures at nano- to micro-structural scales. At the microscale, the extreme volume change may cause delamination of the electrode from the current collector and particle isolation from the conductive carbon network in the laminate.[3] At the nanoscale, the cyclical loads from expansion and contraction of the Si particles may crack the solid electrolyte interphase (SEI) and expose the reactive Si surfaces, enabling further electrolyte reduction.[4,5] During extended cycling, the continuous decomposition of the electrolyte causes solvent and Li inventory loss (i.e., irreversible loss of capacity), as well as the accumulation of SEI fragments (i.e., porosity reduction) and loss of electrical percolation (i.e., increased resistance) in the Si electrode.[6–8]

A substantial body of research has been published on the cycle life of the Si-based batteries. For example, nano-engineering of Si particles has been shown to reduce mechanical strain and improve capacity retention over hundreds of cycles.[9–12] Modified polymer binders such as poly (acrylic acid sodium)-grafted carboxymethyl cellulose, crosslinked poly(acrylic acid) (PAA)-polyvinyl alcohol (PVA), and boronic crosslinked polysaccharides, have been reported to enhance mechanical integrity and cycling stability by providing a strong adhesive network for the electrode.[13–15] One popular method of achieving relatively stable SEI for Si cycling is its in-situ modification with electrolyte additives. Additives such as fluoroethylene carbonate (FEC), vinylene carbonate (VC), and lithium difluorooxalatoborate (LiFOB) has improved capacity retention in Si anode-based cells by 12–14% over baseline electrolyte.[16] FEC and VC have been reported to form conformal SEIs purported to suppress electrolyte consumption.[17,18] Additionally, incorporating N-carboxyanhydrides as an additive enhances the capacity retention of Si||NMC532 cell, allowing them to undergo approximately 100 to 190 more cycles compared to the baseline electrolyte before reaching a state of health of 60%.[19] A ternary mixture consisting of FEC, VC, and nitrile-functionalized silanes (TEOSCN) has been reported to form a thin, passivating SEI. The TEOSCN-boosted SEI significantly enhances retention, retaining 58% of the initial capacity after 417 cycles compared to 42% in the absence of TEOSCN.[20] Among the additives, FEC has garnered the most attention as cells with FEC in the electrolyte consistently display improved lithiation of Si and markedly increased capacity retention relative to the baseline electrolyte without FEC.[21,22] Moreover, the SEI formed with FEC has higher ionic and electronic conductivity compared to the SEI formed with VC, enabling improved cycling over a wide temperature range of −5 to 60 °C.[23]

Similar to cycling, batteries may lose capacity while in storage (at rest) due to thermodynamically spontaneous chemical reactions between the electrodes and electrolyte, a phenomenon quantified by the calendar life metric.[24,25] In electric vehicle and grid storage applications, batteries experience both active use (cycle) and rest (calendar) aging. Factors such as expansion-induced damage to the Si electrode and SEI fracture that dominate during cycling are absent in calendar aging, i.e., the two degradation modes have largely different mechanisms.[26] Thus, success in cycle life does not necessarily translate into long calendar life. Calendar aging of Li-Ion cells is reported to be exacerbated with Si-based anodes [24] due to the high reactivity of the Si surface.[25] The rate of parasitic reactions is reported to be 5–15 times higher on the Si anode surface compared to the cathode surface at open circuit voltage (OCV).[27] Storage conditions (temperature, state-of-charge) and electrolyte formulation strongly influence calendar aging. For instance, Rodrigues et al.[28] found that storing pouch cells with Si-Gr (15:73 wt%) composite anode for 4 years in the charged state results in a 13% reduction in state of health (SOH), whereas storing in the discharged state exhibit negligible decrease in SOH. On the other hand, using Si-rich anodes (70 wt% SiO, 20% LiPAA, 10% C45 carbon tested in SiO||NCM523+ $Li_5FeO_4$ for prelithiation), Lu et al.[29] found that cells containing 10 wt% FEC show ~0.2% capacity loss per day regardless of active cycling or extended rest. Fundamental scanning electrochemical microscopy (SECM) study of Si in 1.2M $LiPF_6$ in EC:EMC (3:7 w/w, no additive) revealed an order of magnitude greater chemical reactivity of the delithiated Si thin film compared to the lithiated.[30] This finding suggests the passivation of the SEI at rest might be more compromised in the delithiated state of Si, which does not straightforwardly match with the above findings of greater capacity loss whence the Si is the lithiated state during calendar aging. One can speculate that whence the Si is in the lithiated state before calendar storage the resulting loss of Li inventory (LLI) as the Si reacts with the electrolyte (milder it might be) leads to irreversible capacity loss. With minimal Li present in the delithiated state, less cyclable Li inventory loss occurs and thereby less irreversible capacity loss.

Electrolyte formulation, particularly the choice of solvents and additives, affects the calendar life of Si anodes in manners not yet completely understood. The SEI on Si formed from conventional carbonate solvent + $LiPF_6$ electrolytes does not remain inert during OCV. SEI dissolution reportedly onsets within the first 45 hours of OCV and this rate predictably increases with temperature.[31] Propylene carbonate (PC) + FEC electrolytes promote dense LiF-rich SEI and has shown positive effects on calendar life, whereas EC+FEC forms a thick SEI and an increase in impedance is measured while cells are stored at 60°C.[32] Moreover, FEC-free localized high-concentration electrolytes have been shown to suppress Si–cathode crosstalk, minimizing impedance growth, and enhancing calendar life compared to FEC containing EC: EMC: DMC (vol 1:1:1) electrolyte.[33] Assessment of several additives against Si-rich anodes by Verma et al.[34] using potentiostatic hold accelerated aging protocols in Si||$LiFePO_4$ cells indicates EC free FEC:EMC may improve the calendar life over reference EC:EMC:FEC electrolyte by 100 to 300%. Verma et al.[35] also report that a non-fluorinated electrolyte system (EC:EMC + 0.7 M LiBOB) may offer a calendar life 1.6 times longer compared to fluorinated EC:EMC + 10 wt% FEC with composite Si-Gr anodes, albeit the presence of LiBOB results in greater impedance. Given the diverse findings on how electrolyte formulation affects calendar life, it remains crucial to refine our understanding of SEI evolution and eventual passivation or lack thereof during aging. The electrochemical reduction products during the initial SEI formation by cycling may not have the same properties as deposits formed through reactivity during calendar aging.

The impact of major additives such as FEC on calendar aging may differ fundamentally from its well-established role in cycle life. Previous investigation using Si thin-film||$LiFePO_4$ cells clarified that the deposit of polymeric carbonate species from FEC is predominantly catalyzed by partially delithiated Si under calendar storage conditions. The reinforcement of the SEI by these polymer fragments during OCV is credited to decrease post-aging capacity loss (by 43% in our thin film studies) over the baseline electrolyte where no polymer fragments are detected in the SEI after aging.[36] However, thin-film electrodes do not capture the heterogeneous environments of porous Si anodes in practical cells where binders, conductive additives, and microstructural porosity might strongly influence the SEI evolution.

This critical gap motivates the present study. We perform a comparative investigation of the calendar aging in full cells with high loading porous Si anodes and EC:EMC (3:7 w/w) + 1.2 M $LiPF_6$ electrolyte, with and without 10 wt% FEC. $LiFePO_4$ (LFP) is selected as the cathode counter-electrode because its flat voltage profile (< 3.5 V vs. $Li/Li^+$) lies safely within the carbonate electrolyte stability window. Minimizing cathode-driven parasitic processes permits the reasonable attribution of most degradation to the Si anode under investigation.[37–39] The present calendar aging study finds that FEC markedly reduces irreversible capacity loss from 7.47% to only 0.57% following 192 hours (8 days) of aging; data model extrapolation predicts a 13-fold longer calendar life compared to FEC-free cells.

Electrochemical impedance spectroscopy (EIS) paired with distribution of relaxation times (DRT) analysis of symmetric cells (Si||Si) using pre-conditioned Si electrodes further reveals significant growth in both SEI ($\Delta R_{SEI}$ = 10.81 Ω) and charge-transfer resistances ($\Delta R_{ct}$ = 18.85 Ω) during ~2 months of aging in the absence of FEC. On the other hand, Si aged in symmetric cells with FEC as additive exhibit approximately 3 – 4.7 times lower impedance changes ($\Delta R_{SEI}$ = 3.37 Ω and $\Delta R_{ct}$ = 3.96 Ω). Pre- versus post-aging FTIR and XPS identify deposition of polycarbonate species only in Si electrodes aged in presence of FEC (and not after electrochemical formation cycles), establishing that FEC uniquely undergoes self-polymerization during the calendar aging to form a more protective SEI. These polymeric films suppress parasitic reactions during idle shelf-storage, as supported by a lower resistance growth exponent (0.49 for 10 wt% FEC vs. 0.62 for 0 wt% FEC) obtained from power law fitting of impedance versus time. When Si electrodes aged for 63 days in symmetric cells are reassembled into full cells and cycled at C/10, 10 wt% FEC cell exhibits an enhanced average coulombic efficiency (99.7% with 10 wt% FEC vs. 95.3% with 0 wt% FEC). Upon five additional cycles at C/3 for these cells with harvested aged Si electrodes, 10 wt% FEC cells retain 95.6% of their initial C/3 capacity, whereas 0 wt% FEC cells retain only 65%. These insights suggest for the first time that a potential approach to boosting the calendar life of Si-based cells may be to utilize advanced self-polymerizing additives in combination with the established aging step during industrial cell pre-conditioning for in situ generation of a thin reinforced and passivating polymer-rich SEI.

## 2. EXPERIMENTAL

2.1. **Cell preparation.** All electrode laminates were provided courtesy of the Cell Analysis, Modeling, and Prototyping (CAMP) Facility at Argonne National Laboratory. The high-loading Si anode contains 80 wt.% Si produced by Oak Ridge National Laboratory (ORNL), 10 wt.% C45 carbon (Timcal$^{TM}$) and 10 wt.% p84 Polyimide binder. The LFP cathode contains 92 wt.% LFP (Johnson Matthey), 3.9 wt.% C45 carbon, 0.1 wt.% single-walled carbon nanotubes (SWCNT, Tuball$^{TM}$) and 4.0 wt.% PVDF binder (Solvay 5130$^{TM}$). Electrodes and separators were vacuum-dried respectively at 110°C and 70°C overnight prior to cell assembly. Two electrolyte formulations were investigated: 1.2 M $LiPF_6$ in EC: EMC (3:7 w/w) as baseline, and baseline + 10 wt% FEC additive. Electrodes and separator were punched to 14 and 16 mm diameter disks, respectively and assembled in CR2032-type coin cells (Si || 40 µL Electrolyte in Celgard 2325 || LFP). **Table 1** provides the nomenclature used for samples using each of the two electrolyte formulations.

**Table 1.** Nomenclature of Si||LFP full cells according to the electrolyte used.

| Electrolyte | Nomenclature |
|---|---|
| EC:EMC 3:7 w/w + 1.2M $LiPF_6$ | 0 wt% FEC |
| (EC:EMC 3:7 w/w + 1.2M $LiPF_6$) + 10 wt% FEC | 10 wt% FEC |

2.2. **Electrochemical characterization.** All electrochemical cycling tests were performed with a Neware battery testing system (Version 8.0) with the cells inside a temperature-controlled chamber set at 30°C. It is worth reiterating that investigative assessment is centered on the Si anode aging in presence and absence of FEC. The galvanostatic potential versus time of the LFP electrode in an LFP||Li half-cell is shown in **Figure S1** displaying the anticipated flat profile between 3.4 – 3.46 V. Cai et al.[27] affirmed the spectator status of the LFP cathode as compared to the Si electrode in most aging processes.

2.2.1. **Cell aging.** A protocol for gauging calendar aging involves monitoring the cell at open circuit (resting state OCV) for long periods of time in the charged state (lithiated Si) after the formation cycles. Following this OCV period, a reference performance test is performed to gauge irreversible losses in capacity resulting from the time at rest. As parasitic reactions consume Li inventory and cause possible particle and electrode degradation, a negative (post-OCV – pre-OCV) capacity difference is straightforwardly a proxy measure of irreversible processes.[27]

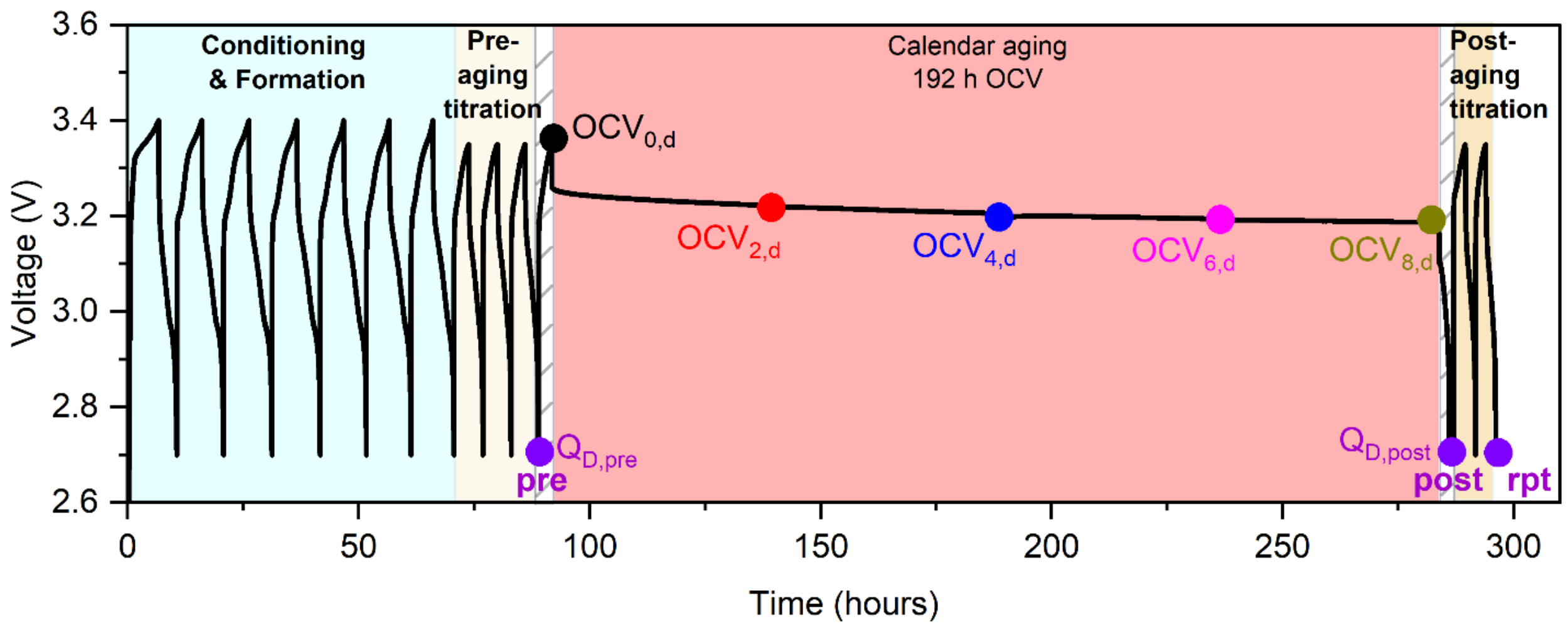

**Figure 1.** Electrochemical protocols used for analysis of the calendar aging of Si||LFP cells.

**Figure 1** shows the protocol used to investigate the calendar life of Si||LFP cells with and without FEC in the electrolyte. The protocols contain four major regions. The first region is an initial conditioning and formation cycling consisting of seven charge/discharge cycles between 2.7 – 3.4 V at C/10. Following the conditioning cycles is a pre-aging titration region, consisting of three additional cycles between 2.7 – 3.35 V at C/10 current; the last discharge cycle (marked

"pre" in **Figure 1**) capacity ($Q_{D, pre}$) is used as the pre-aging cell capacity. The third region denoted "Calendar aging" consists of first charging the cell to 3.35 V, then leaving the cell at OCV (zero current) for 192 hours (8 days). The region ends with a galvanostatic (CC) discharge at C/10 to 2.7 V followed by potentiostatic hold (CV) at 2.7 V till current reached C/50. The post OCV-aging capacity is measured at the end of 11$^{th}$ discharge (CC + CV) and denoted by $Q_{D, post}$. The post-aging titration is the final region, and it consists of two reference performance tests (RPT) charge/discharge cycles at C/10 between 2.7 – 3.35 V; the second discharge during this RPT ($Q_{D, rpt}$) measures the cell capacity post-aging. **Table 2** provides the equations used for the assessment of aging in Si||LFP cells.

**Table 1.** Equations used for quantification of the Si-based cell aging behavior. *t* is time (day).

| | |
|---|---|
| $\Delta Q\ (\%) = \dfrac{(Q_{D,rpt} - Q_{D,pre})}{Q_{D,pre}} \times 100$ | …… Eq. 1 |
| $Q_{irrev}\ (\%) = \dfrac{(Q_{D,pre} - Q_{D,post})}{Q_{D,pre}} \times 100$ | …… Eq. 2 |
| $a = \dfrac{Q_{irrev}\ (\%)}{(t_{192\ h})^p}$ | …… Eq. 3 |
| $Relative\ capacity\ (\%) = 100 - (a \cdot t^p)$ | …… Eq. 4 |

Irreversible capacity loss ($Q_{irrev}$) arising from SEI growth is determined from the difference between the discharge capacity measured before aging ($Q_{D, pre}$) and after aging ($Q_{D, post}$), as expressed in Eq. 2. The evolution of $Q_{irrev}$ is commonly described by a power-law dependence on time (Eq 3), where $a$ is a prefactor and $p$ is the time exponent that characterizes the SEI growth kinetics. In voltage-hold aging experiments, both $a$ and $p$ can be determined by fitting $Q_{irrev}$ as a function of time.[35] However, for OCV calendar aging tests which are more representative of cells in storage, only one $Q_{irrev}$ data point can be calculated (after minus before aging), and thus $p$ cannot be independently extracted. Physics-based models for diffusion-limited SEI growth predict $p = 0.5$ whereby the rate of capacity loss decreases over time as the SEI layer thickens and transport of reactive species through the SEI becomes restricted.[40,41] For highly passivating SEI, values of $p < 0.5$ have been suggested, while for reaction-limited SEI growth, $p = 1$.[35] Since calendar-life experiments are typically of short duration and SEI growth in Si may be accelerated by processes such as temperature, or SEI dissolution with exposure of reactive Si surfaces,[25,33,42] adopting $p = 0.5$ may lead to an overestimation of calendar life of Si (**Figure S2**). Here, the reaction-limited case ($p = 1$) is adopted, which represents an accelerated aging scenario in which $Q_{irrev}$ increases linearly with time. This assumption is conservative and reflects extreme conditions of calendar aging, under which SEI growth is sustained by continuous surface reactions without transport limitation. Calendar life, defined as the time for a lithium-ion battery's capacity to decline by a specified fraction, is computed from Eq. 4. Application specific thresholds are commonly used: 20% capacity loss for electric vehicle (EV) applications and 50% for aircraft applications.[43,44] In this work, we evaluate calendar lifetime using the 20% capacity loss criterion appropriate for EVs.

**2.2.2. EIS and DRT analysis.** To monitor changes in impedance during aging, potentiostatic electrochemical impedance spectroscopy (EIS) was performed using a 5 mV amplitude over the frequency range of 100 kHz to 10 mHz. The data points, represented by filled circles in **Figure 1**, mark the time at which EIS measurements were acquired. The impedance spectra collected during the aging are labeled $OCV_{n,d}$ where "n" is the number of days since initiation of the OCV aging.

In addition, symmetric cells Si||Si and LFP||LFP were constructed using both electrolyte formulations to isolate impedance contributions from Si and LFP electrodes. These symmetric cells were assembled using previously cycled electrodes harvested at $OCV_{0,d}$ (pre-conditioned but not yet aged, see **Figure 1**). EIS measurements on these symmetric cells were performed at various time points over ~63 days while idling in the 30°C temperature chamber to differentiate impedance

changes originating from the Si or LFP electrodes. Note that fresh electrolyte and separator were used in assembling the symmetric cells.

Distribution of Relaxation Times (DRT) analysis was performed to resolve individual processes from the EIS data. In the DRT framework, each electrochemical process is modeled as an RC element characterized by a specific time constant (τ), allowing for the separation of processes based on their timescales. The total impedance response is modelled as infinite number of RC elements with continuous distribution of the relaxation time. With the expectation of a finite number of processes in the cell, resistive intensity is expected around a finite set of time constants; the peak area of the distribution function is used to quantify the contribution of each process. The DRT is obtained from the impedance spectrum by inversion of the integral expression in Eq. 5.[45]

$$Z_{DRT}(\omega) = R_o + R_{pol} = R_0 + \int_{-\infty}^{+\infty} \frac{\gamma(ln\tau)}{1+j\omega\tau}\, d(ln\tau) \text{ ..... Eq. 5}$$

The term $R_o$ denotes the DC ohmic resistance. The expression $\gamma(ln\tau)/(1+j\omega\tau)$ quantifies the fractional contribution of processes with relaxation times between τ and τ+dτ. Determining $\gamma(ln\tau)$ from impedance data using Eq. 5 is an intrinsically ill-posed problem, requiring numerical regularization for a stable solution. In this study, the function $\gamma(ln\tau)$ is discretized using Gaussian radial basis functions (RBF), which enables faster convergence and improved approximation for data that are scattered or non-uniform, in contrast to conventional basis functions such as piecewise linear functions (PWL).[45,46] The DRT deconvolution was conducted using the Python-based tool, pyDRT, developed by Ciucci et al.[47] Regularization parameters were selected using the generalized cross-validation (GCV) method, which has been shown to yield the most accurate results on synthetic EIS data.[47]

**2.3. Postmortem Characterization.** To investigate the structural and chemical changes induced by calendar aging, postmortem X-ray photoelectron spectroscopy (XPS), and Fourier Transform

Infrared Spectroscopy (FTIR) were performed on Si electrodes retrieved from Si||Si symmetric cells. Si electrode retrieval was conducted at two time points: (i) at the beginning of aging, i.e., pre-conditioned Si electrodes collected at the 11th charge cycle of the Si||LFP cell ($OCV_{0,d}$ in **Figure 1**), and (ii) after 63 days of continuous calendar aging in symmetric cells (Si||Si). To ensure reliable post-mortem analysis, all electrodes were processed following a standardized retrieval and cleaning protocol. After cell disassembly in the glovebox ($O_2$ and $H_2O$ < 0.1 ppm), the Si electrode is gently soaked in dimethyl carbonate (DMC) for approximately 5 minutes to remove residual electrolyte then left to dry in the inert Argon environment. Chemical changes were analyzed by XPS (Thermo Scientific K-Alpha XPS) with an Al $K_\alpha$ source at a spot size of 400 μm. The Thermo Scientific™ Avantage™ XPS software (v6.6.0 Alpha) was used to fit the XPS data, and peak positions were calibrated assuming the XPS signal for C – C bond at 285 eV in the $C_{1s}$. XPS samples were transferred to the XPS load-lock from the glovebox in an air-free stage to avoid air exposure. FTIR in ATR mode (Thermo Nicolet Nexus 670 equipped with a diamond crystal and KBr detector) was performed from 400 – 4000 $cm^{-1}$ with resolution of 4 $cm^{-1}$ and a sample binning of 200. The sample is sealed in a heat-seal bag for transport to the FTIR where the bag is opened and compressed on the crystal as quickly as possible to minimize air exposure.

## 3. RESULTS

To assess the impact of FEC on the calendar aging of Si||LFP cells, cells are first subjected to 10 formation cycles at a C/10 terminated by an 11th charge at C/10 to 3.35 V. The cells are held at open circuit voltage (OCV) for 192 hours to induce calendar aging. The average charge capacity versus cycle number is shown in **Figure 2a**, with error bars representing one standard deviation from triplicate samples and the dotted line at 11th cycle represents the OCV aging initiation. The drop in capacity observed after seven cycles in both 0 and 10 wt% FEC cells correspond to the

narrowing of the charge voltage window from 2.7 – 3.4 V to 2.7 – 3.35 V. The initial charge capacity of the 10 wt% FEC cells is on average 1.13 times greater than their 0 wt% FEC counterparts. This additional capacity is common and attributable to the preferential reduction of FEC, which occurs at a higher potential than ethylene carbonate (EC), thereby contributing extra capacity early in the formation cycles.[48] Over subsequent cycles, gradual decrease in capacity is observed for the 0 wt% FEC cell foreshadowing cycling losses in the Si (trapped Li, particle isolation after amorphization and cracking, consumption of Li inventory in the SEI formation etc).

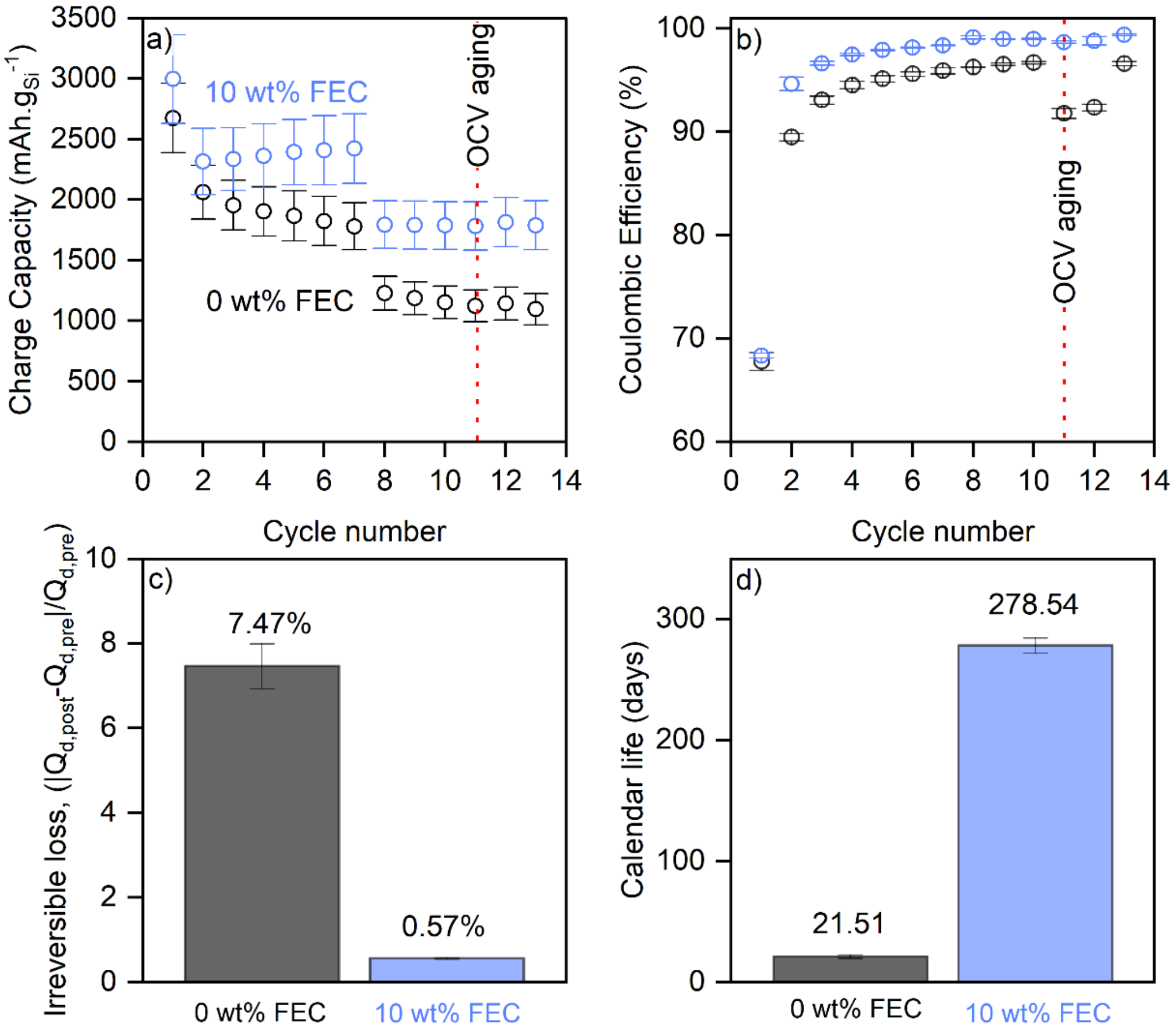

**Figure 2.** (a) Average charge capacity and (b) Coulombic efficiency during cycling pre- and post-OCV aging. Dotted lines in (a) and (b) represent the OCV-aging divide. (c) Irreversible capacity loss, $Q_{irrev}$ of Si||LFP full cells due to the 192 h OCV-aging for 0 wt% FEC and 10 wt% FEC and (d) projected time for the cells to reach 80% of their initial capacity given OCV calendar aging.

In contrast, cells with 10 wt% FEC demonstrate a flat to slight increase in capacity, likely due to a thinner SEI promoting faster $Li^+$ diffusion and charge transfer.[16,49]

During the OCV aging period, a monotonic decline in voltage manifests in both full cells (**Figure S3**). The initial rapid voltage drop is attributed to diffusion-driven lithium concentration homogenization across the Si and LFP electrodes. However, both cells experience monotonic OCV decrease with no clear asymptote after 192 hours (8 days); continued loss of electrode potential is associated with parasitic reactions consuming stored Li. Si gives up $Li^+$ to compensate for electron loss to the electrolyte in parasitic reactions (self-discharge),[27] thereby raising the anode potential which shrinks the cell OCV difference. This voltage decay occurs despite the presumed formation of a stable SEI during the initial cycling phase. Cells without FEC exhibit more pronounced OCV decay, declining from 3.35 V to ~3.06 V after 192 hours. In contrast, cells with 10 wt% FEC retain a higher terminal OCV of ~3.19 V. The slower OCV decay with 10 wt% FEC points to FEC promoting a more robust and passivating SEI that mitigates Si self-delithiation.

Coulombic efficiency (CE) trends are shown in **Figure 2b**. While both 0 wt% and 10 wt% FEC cell groups initially exhibit similar CE, cells with FEC show consistently higher CE during subsequent cycles. Following the aging period starting at cycle 11 (marked by dotted line in **Figure 2b**), a marked drop of CE is observed for the 0 wt% FEC cells while the 10 wt% FEC cells retain most of the CE after the aging. During the subsequent rate performance test (RPT), average CE stabilizes above 99% for 10 wt% FEC cells and approximately 94.5% for FEC-free cells as the discharge and charge capacities come back into better balance over the two RPT cycles. Comparing the cycle efficiencies pre- and post- the 192 h OCV aging in **Figure S4**, it is striking that the mean CE for cells aged under 0 wt% FEC decreases while that of cells aged under 10 wt% FEC increases instead. The mean CE for 10 wt% FEC cells improves from 98.8% to 99.1% across the calendar aging, whereas the mean CE for 0 wt% FEC cells drops from 96.52% to 94.5%. Greater irreversible lithium losses occur from calendar aging with the baseline $LiPF_6$ in EC:EMC electrolyte while the SEI under FEC appears to better suppress parasitic reactions after the 192 hours calendar aging.

To estimate the calendar life of the Si||LFP full cells, first the irreversible capacity loss ($Q_{irrev}$) during OCV aging is calculated using Eq. 2 in **Table 2**. Next, parameter '$a$', slope of $Q_{irrev}(t)$ during the 192 hours OCV, is obtained using Eq. 3. The parameter $p$ is set to 1 as motivated in the experimental section. The relative capacity retention is then projected using Eq. 4 in **Table 2**. The irreversible capacity loss of Si||LFP cells during the 192 hours OCV idling period and the projected time required for the cells to lose 20% of their initial capacity are presented in **Figure 2c** and **Figure 2d**, respectively. Cells without FEC experience a significant irreversible capacity loss of approximately 7.47% over 192 hours of rest, which is consistent with the substantial self-discharge behavior observed in **Figure S3**. In contrast, the inclusion of 10 wt% FEC in the electrolyte formulation dramatically mitigates this effect, resulting in a minimal capacity loss of just 0.57% over the same 192 hours of OCV calendar aging. Based on the observed irreversible loss during aging, the average calendar lifetime is estimated to be 22 days for the FEC-free cells. The addition of 10 wt.% FEC to the electrolyte extends the projected calendar life to 279 days. Overall, the incorporation of FEC into the baseline electrolyte reduces the irreversible capacity loss by approximately 13-fold and the projected calendar lifetime is 13-fold longer.

The processes engendering the loss of capacity during calendar aging, i.e, loss of Li from self-delithiation into additional SEI products, electrical and ionic isolation of Si, and $Li_xSi$ must have an impedance signature. The temporal evolution of impedance associated with SEI dynamics during OCV calendar aging is monitored by EIS coupled with DRT analysis at two-day intervals (indicated in **Figure 1**). Representative Nyquist plots and corresponding DRT spectra are shown in **Figures 3a, b** (0 wt% FEC) and **Figures 3c, d** (10 wt% FEC). DRT enables the deconvolution of impedance contributions from distinct electrochemical processes, including contact resistance, interfacial polarization, charge-transfer resistance, and mass transport.[50] Each process manifests within a range of time constant $\tau$, allowing for peak assignment: SEI processes ($R_{SEI}$) are typically reported within $\tau = 10^{-4}$–$10^{-2}$ seconds, whereas charge transfer ($R_{ct}$) is expected within $\tau = 10^{-2}$–$10^{0}$ seconds.[51,52]

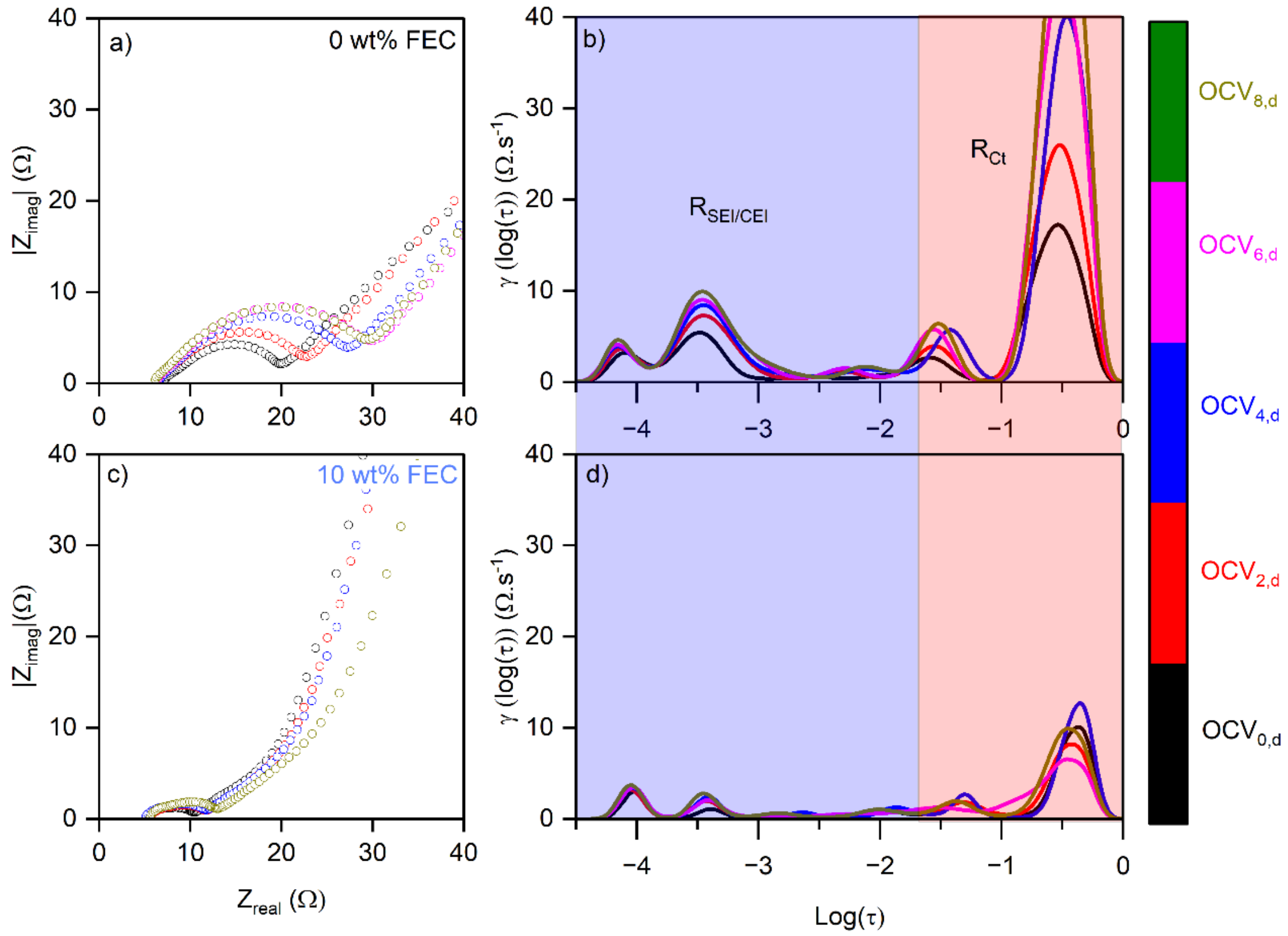

**Figure 3.** Nyquist and corresponding DRT of Si||LFP full cells whence 0 wt% FEC (a, b) versus 10 wt% FEC (c, d) are added to the baseline electrolyte. Data was collected at 2-day intervals during the 8 days (192 h) of OCV aging period.

To disentangle the electrode-specific contributions (either Si or LFP) to the overall Si||LFP cells impedance, symmetric Si||Si and LFP||LFP cells are assembled using electrodes harvested at $OCV_{0,d}$ (see **Figure 1**). The corresponding DRT spectra for these symmetric cells with baseline electrolyte are shown in **Figure S5**. In DRT, the Si anode SEI-related peaks overlap with the cathode–electrolyte interphase (CEI) peaks of LFP, while the $R_{ct}$ peaks of both electrodes fall within the $10^{-2}$–$10^{0}$ sec region (comparing **Figure S5b-ii** to **Figure S5c-ii**). Therefore, in the full cells, peaks within the $10^{-4.5}$–$10^{-2}$ sec domain are assigned to $R_{SEI/CEI}$ and peaks within the $10^{-2}$–$10^{0}$ sec region are attributed to $R_{ct}$ (**Figure 3b, d**).

The Nyquist plots in **Figure 3a** show the semicircle diameter increasing progressively over the 8 days (192 hours) of OCV aging for the 0 wt% FEC cells, indicating continuous interfacial impedance growth and charge transfer resistance buildup. In contrast, cells containing FEC exhibit only minimal impedance growth over the same aging duration (**Figure 3c**). DRT spectra in **Figure 3b** confirms a pronounced increase in peak intensity associated with SEI/CEI and charge-transfer resistance in the 0% FEC cells over the aging. In contrast, the FEC-containing cells display negligible growth in these features, highlighting the stabilizing effect of FEC on interfacial processes. Furthermore, peaks assigned to the SEI/CEI region in the FEC-free cells shift slightly toward longer relaxation times (**Figure 3b**), indicating slower interfacial kinetics and progressive SEI/CEI thickening. In contrast, the FEC-containing cells in **Figure 3d** maintain nearly constant peak positions, suggesting the formation of a more stable SEI in the presence of FEC. To quantify the temporal evolution of SEI/CEI and charge transfer-related impedance, the area under the SEI/CEI and charge transfer-associated DRT peaks are calculated and plotted as a function of aging time in **Figure 4a** and **4b** respectively. The evolution of interfacial resistance during the aging period in **Figure 4a** highlights a stark contrast between cells with and without FEC. For the FEC-free aging, the average $R_{SEI/CEI}$ increases from 4.94 Ω to 8.18 Ω (ΔR = 3.24 Ω) from pre- to post- calendar aging in 8 days, representing a ~66% increase. This pronounced rise reflects significant interfacial changes that explain 7.47% capacity loss across the aging in **Figure 2c**. In contrast, the FEC-containing cells exhibit a markedly smaller increase, with $R_{SEI/CEI}$ rising from 1.65 Ω to 2.19 Ω (ΔR = 0.54 Ω), corresponding to only ~32% increase. The twofold slower impedance growth in the presence of FEC highlights a more stable SEI and/or the formation of a passivation layer early in the aging.

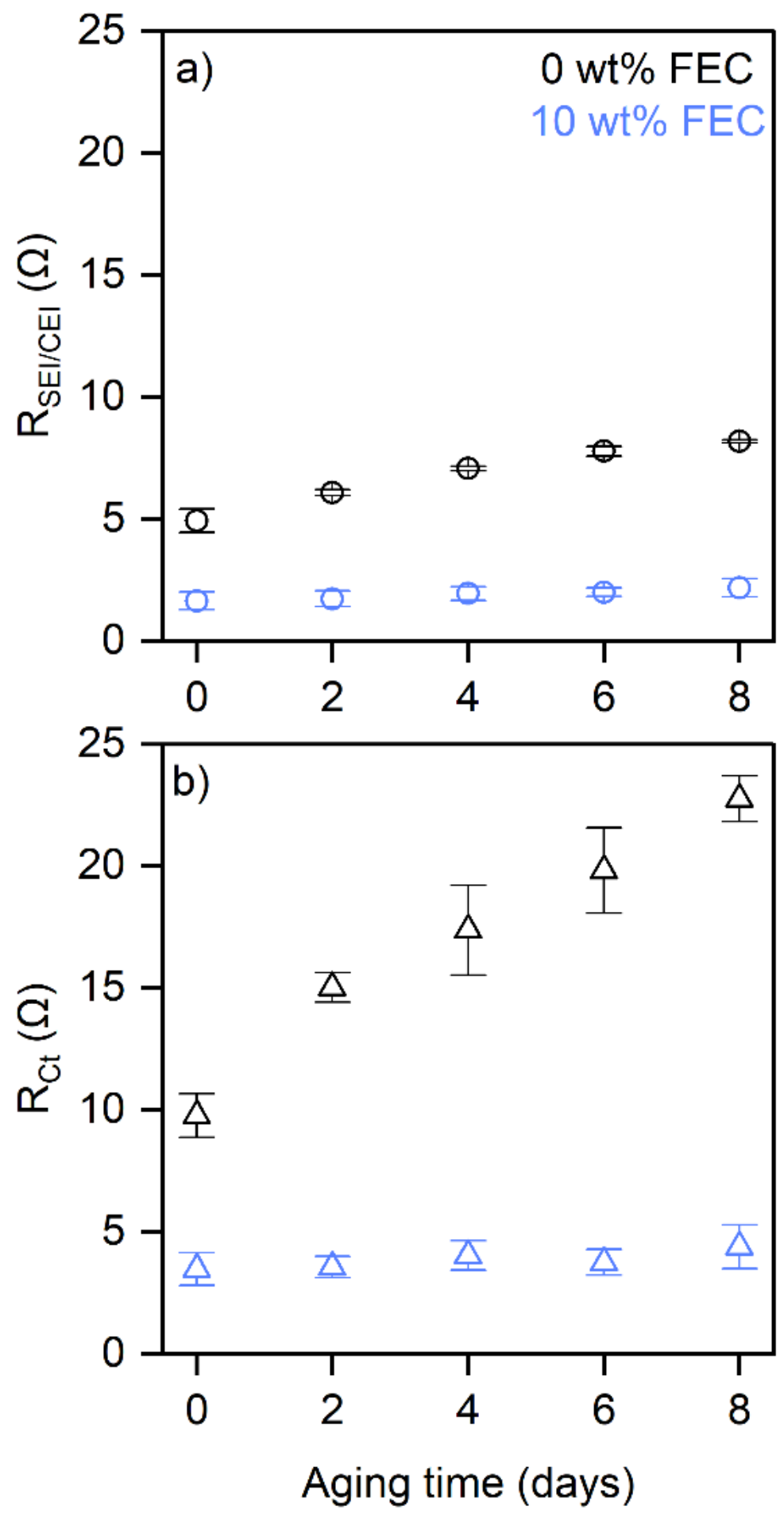

**Figure 4**. Evolution of (a) interfacial SEI and CEI and (b) charge transfer resistance for Si||LFP full cells during the 8 days (192 h) of OCV aging period.

A similar trend is observed for the charge transfer resistance, $R_{ct}$ (**Figure 4b**). The FEC-free cells show a significant 2.3-fold increase in $R_{ct}$, rising from 9.76 Ω to 22.76 Ω. By comparison, the FEC-containing cells exhibit only a modest 1.25-fold rise in $R_{ct}$, from 3.48 Ω to 4.38 Ω. Impedance growth during aging can be ascribed to pore clogging due to SEI formation,[53] gaseous product formation,[25,54] surface transformation of the cathode[54] or side reaction near cathode surface.[27] Comparative peak intensity analysis in **Figure S5** clearly indicates that the dominant impedance contributions in the Si||LFP full cells originate from the Si electrode; substantially higher interfacial and charge-transfer peak intensities observed in the Si||Si symmetric cells (**Figure S5b-ii**) compared to the LFP||LFP counterpart (**Figure S5c-ii**) highlights the critical role of Si in limiting the total interfacial kinetics. However, the symmetric cells in **Figure S5** are examined at the onset of aging. Thus, there exists the possibility of large cathode impedance growth over extended aging durations which is investigated.

Systematic calendar aging EIS studies using Si||Si and LFP||LFP symmetric cells are extended to 63 days (2 months). These cells are assembled using pre-conditioned Si and LFP electrodes harvested at $OCV_{0,d}$ (after $11^{th}$ charge in **Figure 1**); steady-state electrochemistry-derived SEI and CEI are assumed to have formed on these electrodes. Fresh electrolyte are used in the symmetric cells whereby electrodes harvested from formation in 0 wt% or 10 wt% FEC are aged with the same but fresh 0 wt% or 10 wt% FEC electrolyte. **Figure S6a, b** and **Figure S6c, d** present the EIS and corresponding DRT profiles for Si symmetric cells aged in the baseline electrolyte with 0 wt% FEC, and 10 wt% FEC, respectively. It is noted that the symmetric cell records the combined

impedance from both Si electrodes; thus, both x and y axis in the Nyquist plots in **Figure S6** are halved to reflect the contribution of a single Si electrode.

For the FEC-free cells, the Nyquist plots show pronounced impedance growth during the 63 days of aging (**Figure S6a**). The corresponding DRT spectra indicates a progressive shift and increase in SEI-associated peaks, signifying continuous interfacial changes (absent cathode–anode crosstalk effects given the use of symmetric cells). The high reactivity of Si electrodes drives parasitic reactions with the electrolyte, causing substantial impedance growth. In contrast, cells containing FEC exhibit smaller initial semicircle diameters in the Nyquist plot with more subdued growth over the 63 days (**Figure S6c**).

Quantitative evaluation of $R_{SEI}$ and $R_{ct}$ during symmetric cells aging is performed by integrating the corresponding DRT peaks and shown in **Figure 5a** and **Figure 5b**, respectively. **Figure 5a** shows that $R_{SEI}$ for the Si electrode aged in 0 wt% FEC electrolyte increases from 2.43 Ω at day 0 to 4.96 Ω within the first week, and further to 13.24 Ω after two months of aging (~5.4-fold increase). The $R_{ct}$ increases from 3.98 Ω at the onset to 22.9 Ω (~5.8-fold). In contrast, Si electrodes aged in 10 wt% FEC show lower initial $R_{SEI}$ (0.88 Ω), which increased to only 4.25 Ω after two months. Although the relative growth (~4.8-fold) indicates ongoing interfacial change during aging in presence of FEC, the absolute resistance remains more than three times lower compared to without FEC. For $R_{ct}$, only a 2.15-fold increase is observed after 63 days versus nearly 6-fold without FEC. Much greater interfacial restructuring and continued SEI thickening manifests when using the baseline 1.2 M $LiPF_6$ EC:EMC (3:7) electrolyte formulation compared to with 10 wt% FEC additive.

Impedance growth due to interfacial film deposition adheres reasonably to a power law versus time.[41,55,56] Thus the growth of Si SEI impedance during aging in symmetric cells is modeled with $R_{SEI} = b \cdot t^m$, where t is aging time (days) and $b$ is a prefactor. The growth exponent $m$ varies from 0 to 1 depending on the rate-limiting process in the SEI growth.[56,57] For graphite electrodes, SEI growth during calendar aging is typically diffusion-controlled, resulting in an exponent of ~0.5 throughout the aging.[55] In contrast, Si-based electrodes exhibit variations in SEI growth mechanism.[25,58] Our experimental data could be roughly separated into two regions of growth exponent $m$ values: one corresponding to the early aging period (0–8 days) and another for the longer term period (22–63 days). Fitting was performed by linearizing the power-law equation through logarithmic transformation $\ln R = m \cdot \ln t + \ln b$ and applying a line fit. The fitting parameters and corresponding $R^2$ values are provided in **Table S1**.

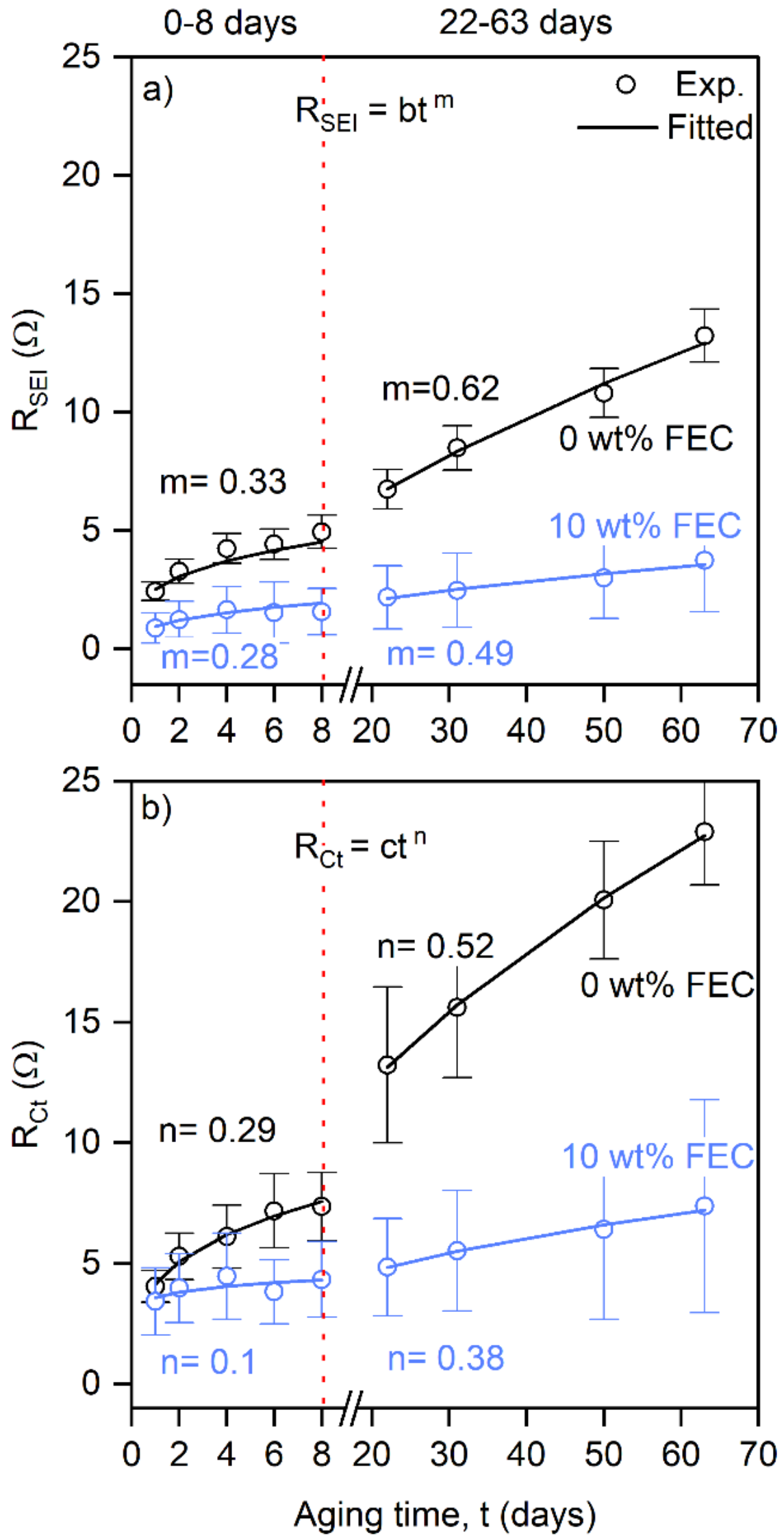

**Figure 5**. Evolution of (a) SEI resistance and (b) Charge transfer resistance of Si electrodes during symmetric cell aging of 63 days. $R_{SEI}$ an $R_{ct}$ are compared between 0 wt% and 10 wt% FEC electrolyte. The resistance values were determined by integration of the corresponding DRT peaks.

During the initial 8 days of aging, the exponent $m$ is slightly higher for the 0 wt% FEC cells compared to the 10 wt% FEC cells (0.33 vs. 0.28). Between 22 and 63 days of aging, $m$ for the 0 wt% FEC cells increases to 0.62, indicating an acceleration of SEI growth with time. This rise in exponent does not represent a purely kinetic regime ($m$ = 1 for reaction-controlled growth), but rather a transition from sub-diffusive and transport-limited growth to a mixed transport–reaction regime. In the absence of FEC, the SEI becomes progressively less passivating, possibly due to dissolution or cracking that exposes fresh electrode surface and promotes continued SEI formation. In contrast, the Si electrodes aged in 10 wt% maintain a growth exponent of ~0.49 in the long-term, consistent with diffusion-limited SEI growth as seen with graphite anodes; continued reaction between Si and the electrolyte requires diffusion of molecules through a persistent and passivating interphase film.

SEI growth also influences charge-transfer resistance ($R_{ct}$), as pore clogging caused by SEI thickening hinders the movement of $Li^+$ ions and electrons within the electrode structure.[42,59] To capture this behavior, a similar power-law $R_{ct} = c \cdot t^n$ is fitted. Here, short-term aging (0–8 days) yields an exponent $n$ of 0.28 for the 0 wt% FEC and 0.10 for the 10 wt% FEC cells. During the extended aging period (22–63 days), the exponent increases to 0.52 for the 0 wt% FEC cells, reflecting a stronger time dependence of $R_{ct}$. Conversely, $n$ increases only to 0.38 in the long-term period of aging in presence of FEC. Here, the significantly lower growth exponents for charge transfer in presence of FEC indicates that the interphase thus formed is not only comparatively thinner and permissive to charge transfer but reduces pore clogging by reducing accumulation of detached SEI fragments. Collectively, the data systematically reinforces the above findings in full cells that FEC (in 1.2 M $LiPF_6$ in EC:EMC at least) promotes the formation of SEI films during calendar aging that significantly diminishes further reactivity of the Si surface with the electrolyte.

Nyquist plot in **Figure S7** for symmetric LFP||LFP aging exhibit little to no impedance growth in both electrolyte formulation compared to the Si||Si cells (**Figure S6**), confirming the general inertness of LFP cathode during extended aging.

Postmortem surface analyses by FTIR and XPS are conducted on Si electrodes retrieved immediately after formation (pre-aging, at $OCV_{0,d}$ in **Figure 1**) and after 63 days of calendar aging in symmetric Si||Si cells ($OCV_{63,d}$). The pre-aging sample was retrieved within minutes of termination of the conditioning cycles to minimize unplanned calendar aging. Following cell disassembly in the Argon-filled glovebox, electrodes are gently rinsed with DMC and left to dry. Care is taken to minimize air exposure during sample transfers as described in the experimental section.

Before and after the calendar aging, the FTIR spectral features of the Si treated without FEC show no obvious differences (**Figure 6a**). The dominant absorption bands observed in the 1220–1635 $cm^{-1}$ range are attributed to C–O and C=O stretching vibrations, corresponding to the formation of $Li_2CO_3$ and lithium ethylene dicarbonate (LiEDC)—common SEI components derived from the carbonate electrolyte decomposition.[60–64] However, prior AFM morphological studies using model Si thin films showed major morphological changes of the FEC-free SEI accompanying loss of capacity.[60] In essence, the SEI deposits during formation have similar overall chemistry as the additional decomposition products generated during OCV calendar aging under 0 wt% FEC in the 1.2 M $LiPF_6$ EC:EMC.

In contrast, new spectral features within the 1780–1800 $cm^{-1}$ range, commonly attributed to polycarbonate species,[22,64,65] emerge in the 10 wt% FEC electrodes only after the calendar aging (**Figure 6b**, $OCV_{63,d}$). The pre-aging sample which underwent 11 formation cycles in presence of FEC (disassembled and rinsed with DMC with minutes of termination to avoid unplanned calendar aging) does not show these 1780–1800 $cm^{-1}$ features (**Figure 6b,** $OCV_{0,d}$). These results agree with our previous report using Si model thin films to show self-polymerization of FEC in contact with the Si surface, especially after partial delithiation, and the lack of polycarbonate features when idle calendar aging is avoided by quick disassembly and rinsing.[36] Despite starting with lithiated Si electrodes in the symmetric cells, partially delithiated phases emerge from Si self-delithiation as detected by OCV decay in **Figure S3** and directly measured by the $Si_{2p}$ XPS in **Figure 6**.

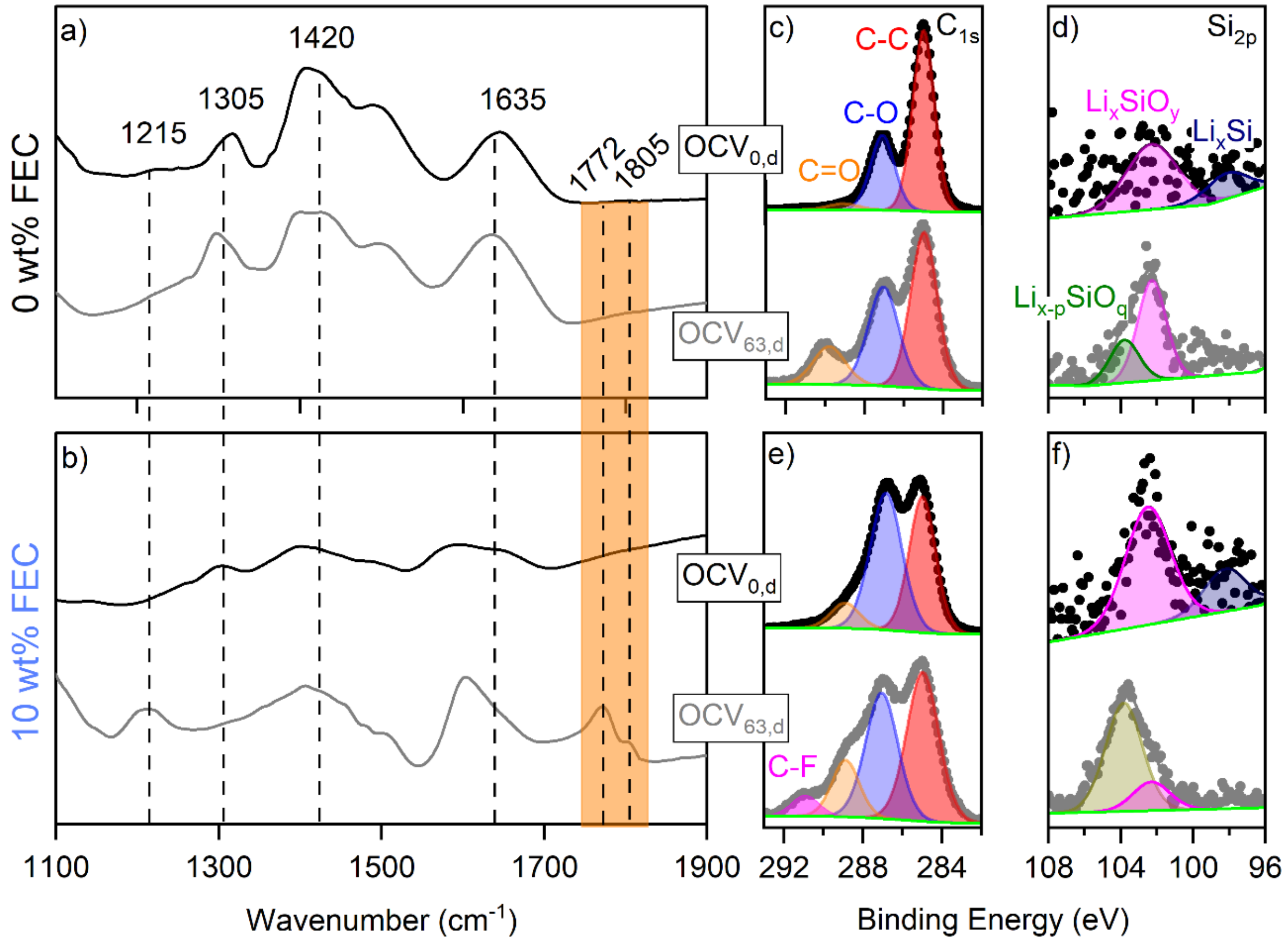

**Figure 6**. FTIR and XPS of Si electrodes for 0 wt% and 10 wt% FEC electrolytes. Panels (a,b) show FTIR spectra. Panels (c,d) present the $C_{1s}$ and $Si_{2p}$ spectra for the 0 wt% FEC aged Si electrode, and panels (e,f) show the corresponding spectra for the 10 wt% FEC aged Si electrode. Spectrums are compared before the aging onset and after 63 days of aging.

To further evaluate lithiation behavior and verify the formation of polymeric species associated with FEC decomposition, XPS is performed. The binding energy associated with different bonds is provided in supporting information (**Table S2**). The resulting high-resolution spectra of Si electrode aged in both FEC-free and 10 wt% FEC cells (**Figure 6c–f**) provide direct evidence of SEI evolution and confirm the presence of polymerization-derived products during calendar aging.

On Si electrodes aged in the 0 wt% FEC electrolyte, the $C_{1s}$ spectrum displays dominant contribution from C–C bonds, accompanied by a consistent C–O signal observed at both 0 and 63 days of aging in symmetric cells. Upon extended aging, the C=O XPS peak intensifies, indicating increased electrolyte decomposition and the accumulation of carbonate-derived species such as LiEDC and $Li_2CO_3$ in the SEI. This interpretation is further supported by the strong $Li_2CO_3$ signature evident in the $O_{1s}$ (C=O) and $Li_{1s}$ spectra (Li–O) after aging (**Figure S8a, b**). Moreover, increases in the Li–F signal intensity within both the $Li_{1s}$ and $F_{1s}$ spectra for $OCV_{63,d}$ electrode (**Figure S8b, c**) points to enhanced salt decomposition and deposition on the SEI during the idle aging period. Analysis of the $Si_{2p}$ spectrum in **Figure 6d** prior to aging shows two distinct features, corresponding to lithium silicide ($Li_xSi$) and lithium silicate ($Li_xSiO_y$). However, after 63 days of aging, the $Li_xSi$ peak is no longer detectable. This disappearance is likely attributable to self-delithiation of Si and/or coverage by decomposition products. Concomitantly, a new feature emerges at higher binding energy, which is assigned to a lithium-deficient silicate phase ($Li_{x-p}SiO_q$), reflecting a reduction in lithium content within the $SiO_x$ matrix.[66–68]

In contrast, for Si electrodes aged in electrolyte with 10 wt% FEC ($OCV_{63,d}$), the $C_{1s}$ spectrum in **Figure 6e** displays an additional feature at 291.2 eV, characteristic of C–F bond.[69,70] This peak is absent from the pre-aging $OCV_{0,d}$ sample retrieved right after 11 electrochemical conditioning cycles in Si||LFP cells. This C–F bond in the $OCV_{63,d}$ sample is attributed to polymeric species derived from FEC self-polymerization at the electrode surface similar to what observed in our previous Si thin film study.[36] The presence of polycarbonate-type species is further substantiated by the $O_{1s}$ spectrum at high binding energy of ~534.5 eV (**Figure S8d**) and agrees with complementary IR spectroscopy data. Examination of the $Si_{2p}$ spectrum in **Figure 6f** for FEC-aged samples shows an initial dominance of lithium-rich silicate ($Li_xSiO_y$), which shifts to higher binding energy after aging, consistent with the formation of lithium-deficient silicate ($Li_{x-p}SiO_q$). The $Li_xSi$ feature becomes undetectable after aging. While the $Li_{1s}$ and $F_{1s}$ spectra of FEC-aged electrodes also show an increase in Li–F intensity (**Figure S8d and S8e**), albeit less pronounced compared to the FEC-free system.

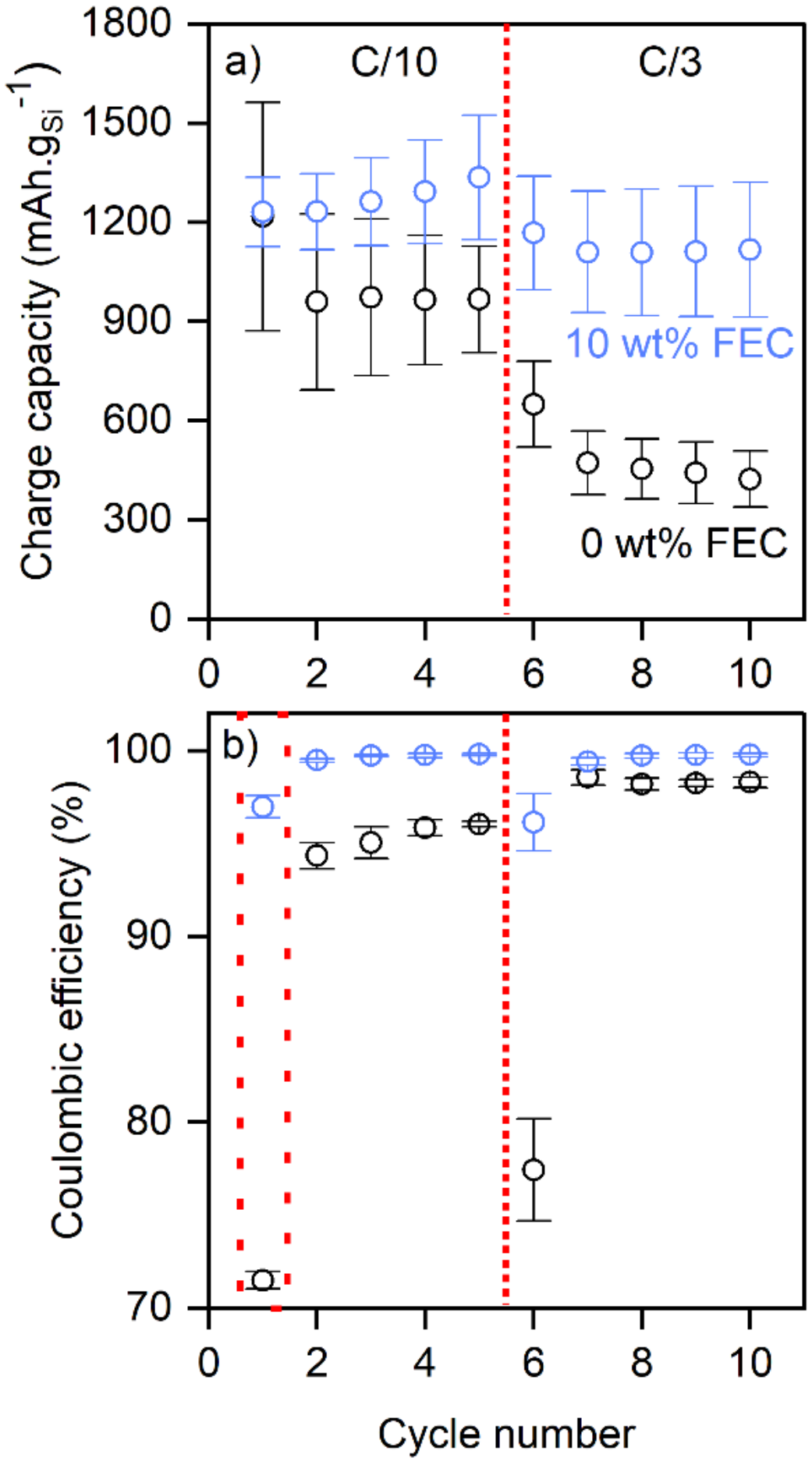

**Figure 7**. (a) Charge capacity and (b) Coulombic efficiency of 0 wt% FEC vs 10 wt% FEC Si||LFP cells using harvested Si electrodes aged 63 days in Si||Si symmetric cells. All other cell components except the Si anode are fresh. Error bars are from triplicate measurements. All the cells are cycled between 2.7–3.35V and capacity is normalized to Si mass.

Following the aging of the Si electrodes in symmetric cells for 63 days, their electrochemical performance is re-evaluated by assembling Si||LFP full cells with the aged anodes. In these cells, fresh electrolyte (with 0 wt% or 10 wt% FEC added), LFP cathodes, and separators are used to ensure that the observed effects originate solely from the aged Si anodes. The cells are subjected to five cycles between 2.7–3.35V at C/10, followed by five cycles at C/3 to assess rate capability (**Figure 7**). Voltage profile of the first two cycle for the assembled cells is provided in **Figure S9** and compared with the voltage profile of the cell pre-aged Si||LFP cells (cells where the Si undergoes only continuous formation and cycling and no calendar aging) from cycle 8-10 during pre-aging titration step.

The initial charge capacities at C/10 of both 0 wt% and 10 wt% FEC cells with 63-days aged Si are comparable within one sigma error at ~1230 $mAh \cdot g^{-1}$ (**Figure 7a**) although the cells with FEC tend to exhibit greater average capacities. Relative to the capacities obtained prior to aging, both cell groups exhibit a ~25-30% reduction (**Figure S9**), albeit the Si electrodes aged with 10 wt% FEC provide higher capacities ca. 1300 $mAh \cdot g^{-1}$ versus ca. 1000 $mAh \cdot g^{-1}$ with 0 wt% FEC. This loss is attributed to less active material availability, likely arising from Si particle isolation, partial dissolution, and lithium loss during aging.[71] FEC-free cells exhibit a rapid decline beginning in the second cycle while cells with 10 wt% FEC maintain more stable performance. The additional ~28% irreversible capacity observed in the 1st cycle of the FEC-free cells is attributed to continued SEI growth. This is further corroborated by the CE in **Figure 7b** and longer charging voltage plateau near 3.21V during 1st lithiation (**Figure S9a**). The SEI constructed on the Si electrodes during the preceding aging in 10 wt% FEC continues to passivate the surface after re-assembly with fresh electrolyte resulting in a 1st cycle CE of ~97%, unlike the SEI from cells aged in 0 wt% FEC where the 1st cycle CE is ~71%. Reduced CE arises from $Li^+$ consumption in electrochemical SEI formation during charging. At the C/10 rate, the 0 wt% FEC cells stabilize at ~965 $mAh \cdot g^{-1}$ after the first five cycles, whereas the 10 wt% FEC cells show an ~8% increase in capacity (on average) reaching 1335 $mAh \cdot g^{-1}$. Increase in capacity over cycles suggests improving active material utilization; the presence of polymer-rich SEI post aging in electrolyte with FEC does not appear to hinder $Li^+$ transport. Instead, the FEC-derived SEI from aging appears facilitate more effective utilization of the active Si.

Upon transitioning to C/3 rate, drop in capacity from kinetic effects is observed in both cell groups. Cells with 0 wt% FEC baseline electrolytes retain only ~65% of their initial C/3 capacity at 10th cycle, whereas the 10 wt% FEC cells maintain ~95.6% retention, highlighting superior rate capability post aging. CE analysis in **Figure 7b** further reinforces this trend: for the FEC-free cells, CE gradually increases to ~96% after five C/10 cycles and then reaches to ~98.3% following the additional 5 cycles at C/3. In contrast, cells using 10 wt% FEC as additive rapidly achieve >99.7% CE after the first C/10 cycle and maintain this value throughout, indicating highly reversible cycling and minimal ongoing SEI growth.

Collectively, these results demonstrate that the polymeric SEI species derived from aging in presence of FEC not only mitigate parasitic $Li^+$ consumption but also enhance electrochemical performance under both slow and fast cycling conditions. One possible counterargument is that the high CE for 10 wt% FEC cells observed after using aged electrodes could simply originate from the electrochemically formed SEI prior to the OCV aging, rather than from polymeric species generated during idle aging. To address this, we examine the C/10 average CE of 10 wt% FEC cells at the pre-aging titration step (cycles 8–10 in **Figure 2b**) and compare it with the average CE obtained from harvested aged electrodes (cycles 2–5 in **Figure 7b**). We exclude the first cycle CE after-aging, since CE on the first cycle after long periods of rest are difficult to interpret (residual capacities, lithium refill after parasitic delithiation, etc.). The average CE is ~98.8% without OCV aging, whereas an elevated CE of ~99.7% is found for harvested Si aged for 63 days in 10 wt% FEC. This contrasts with 0 wt% FEC cells where the CE achieved at C/10 was 96.52% before aging and decreased to 94.34-96.03% after the 63 days of Si electrode aging. The increased CE strongly affirms that the polymeric species formed during the calendar aging in presence of FEC contribute to improved SEI passivation and thereby enhanced reversibility.

## 4. CONCLUSION

A comparative study on the calendar aging of silicon (Si)-based lithium-ion batteries is conducted in the absence and presence of fluoroethylene carbonate (FEC) as an electrolyte additive in EC: EMC (3:7 w/w) + 1.2 M $LiPF_6$. After formation cycles, Si||$LiFePO_4$ full cells are stored at open-circuit voltage (OCV) with the Si electrode in the lithiated state for 8 days to probe the solid electrolyte interphase (SEI) stability under rest conditions.

The full cell calendar aging study resulted in the following key findings:

- FEC-containing cells exhibit markedly lower self-discharge during the 192 hours OCV hold compared to FEC-free cells, reducing irreversible capacity loss from 7.47% to 0.5%.

- The calendar lifetime of Si in full cells is estimated to increase ~13-fold upon the addition of 10 wt% FEC.

To further study calendar aging on the scale of months, pre-conditioned Si electrodes retrieved after $11^{th}$ charge from Si||LFP cell are aged in symmetric cells with or without FEC additive. Characterization by EIS and spectroscopic tools before, during, and after 63 days of aging in Si||Si symmetric cells resulted in the following important insights:

- The emergence of polymeric species as detected by FTIR and XPS on Si aged in electrolyte containing FEC originates from FEC self-polymerization during idle calendar aging and less from reduction during active cycling where very little polycarbonate signals are detected in FTIR and XPS. This result agrees with our prior model system findings on Si thin films where the SEI was probed by FTIR, ToF-SIMS, and XPS. In this study as well as herein, extreme precaution was taken to disassemble the pre-aging electrodes from the cells and rinse with DMC within minutes of completion of cycling to avoid unplanned calendar aging.
- Impedance analysis during aging indicates significant SEI resistance growth ($\Delta R_{SEI} = 10.81\ \Omega$) for Si electrodes aged with no FEC additive, whereas cells aged with 10 wt% FEC exhibits only modest impedance increase ($\Delta R_{SEI} = 3.37\ \Omega$).
- Power law fitting of the interface resistance growth vs. time shows that the SEI with 0 wt% FEC evolves from passivating (exponent < 0.5) to mixed transport-reaction controlled growth (exponent = 0.62) indicating SEI dissolution during 2 months of aging. On the other hand, cells aged under 10 wt% FEC in the electrolyte show slower and diffusion-controlled (exponent ≈ 0.49) SEI impedance growth indicating a more passivating film.

- After 2-month calendar aging of Si electrodes in presence of 10 wt% FEC, cells with the harvested aged Si display elevated coulombic efficiency (>99.7%) compared to Si that only underwent formation in fresh cells (~98.8%). This is attributed to the deposition of polymeric fragments from FEC self-polymerization during idle calendar aging which appears more passivating than the electrochemically-generated SEI alone. Furthermore, cells constructed from harvested Si after aging in 10 wt% FEC electrolyte display superior retention at high rates maintaining 95.6% capacity at the $10^{th}$ C/3 cycle compared to only 65% for harvested Si aged in electrolyte without FEC.

Together, these results demonstrate that Si electrodes aged in the absence of FEC exhibit progressive accumulation of decomposition products derived from both solvent and salt as evidenced through XPS. This layer decreases active material utilization upon resuming cycling. In contrast, the inclusion of FEC leads to the emergence of polymeric species, whose formation likely boosts passivation, suppresses parasitic reactions, and curtails further SEI growth. In turn, lower increases in $R_{SEI}$ and $R_{ct}$ are observed in DRT when compared to the baseline electrolyte formulated without FEC added. Most interestingly, Si electrodes aged in the presence of FEC show signs of improved coulombic efficiency upon resumption of cycling, pointing to a potential advanced pre-conditioning protocol with engineered polymer-generating additives for tailoring the SEI towards cells with greater calendar life. The self-polymerization mechanism of FEC on Si as revealed herein could be a key factor in optimizing the long-term calendar life of silicon-based cells.

ASSOCIATED CONTENT

**Supporting Information**. Calendar aging protocol, Distribution of Relaxation time, Solid electrolyte interphase resistance, Charge transfer resistance, High resolution XPS, coulombic efficiency of 0 wt% FEC and 10 wt% FEC cell.

AUTHOR INFORMATION

**Corresponding Author**

*Koffi P. C. Yao

Email: claver@udel.edu

Department of Mechanical Engineering, University of Delaware, Newark, Delaware 19716, United States of America.

**Author Contributions**

SB performed the experiment, analyzed the data and prepared the manuscript. RJM helped draft the manuscript. KY contributed to supervision, manuscript review and editing, funding acquisition and conceptualization.

**Funding Sources**

This research is supported by the U.S. Department of Energy's Office of Energy Efficiency and Renewable Energy (EERE) under Award Number DE-EE0009185.

ACKNOWLEDGMENT

This work was supported by the U.S. Department of Energy's Office of Energy Efficiency and Renewable Energy (EERE) under Award Number DE-EE0009185. The authors acknowledge the Advanced Materials Characterization Laboratory (AMCL) at the University of Delaware for access to the ATR-FTIR instrument and the Surface Analysis Facility (SAF) for access to the Thermo Scientific K-Alpha XPS system, funded by NSF Award No. 1428149. Additionally, the authors thank the Argonne National Laboratory Cell Analysis, Modeling, and Prototyping (CAMP) Facility and the DOE Silicon Consortium for providing the $LiFePO_4$ and Si electrodes.

**Supporting Information**

# Aging-driven in situ polymerization of FEC additive boosts the calendar-life of silicon anodes via surface passivation enhancement

*Sattajit Barua,[1] Rownak J. Mou,[1] and Koffi P. C. Yao[1*]*

[1]Department of Mechanical Engineering, University of Delaware, Newark, Delaware 19716, United States of America

**Corresponding Author**

[*]Email: claver@udel.edu

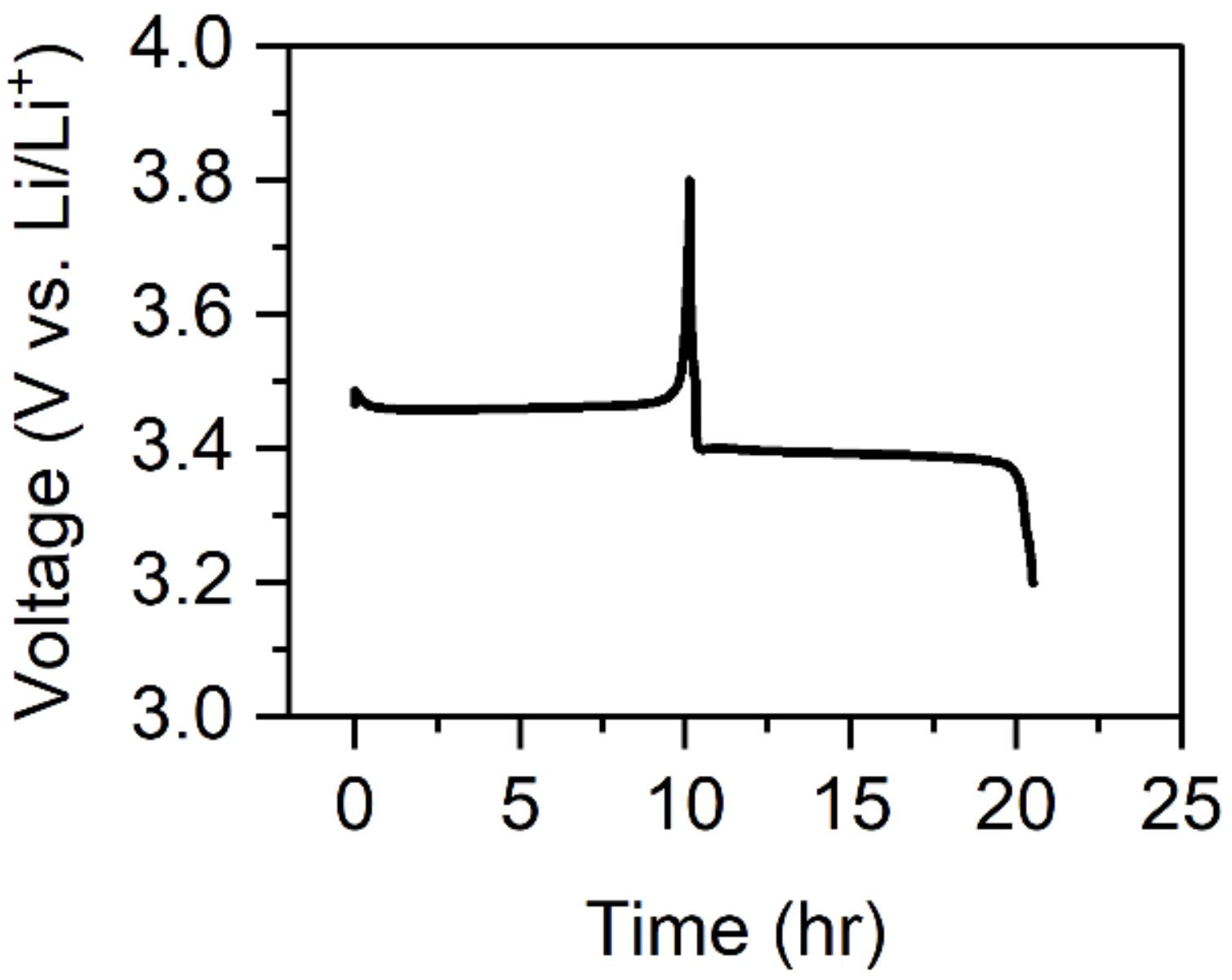

**Figure S8.** Charge and discharge profile of LFP||Li half-cell. The cell is cycled at a current rate of C/10 between 3.2 – 3.8 V.

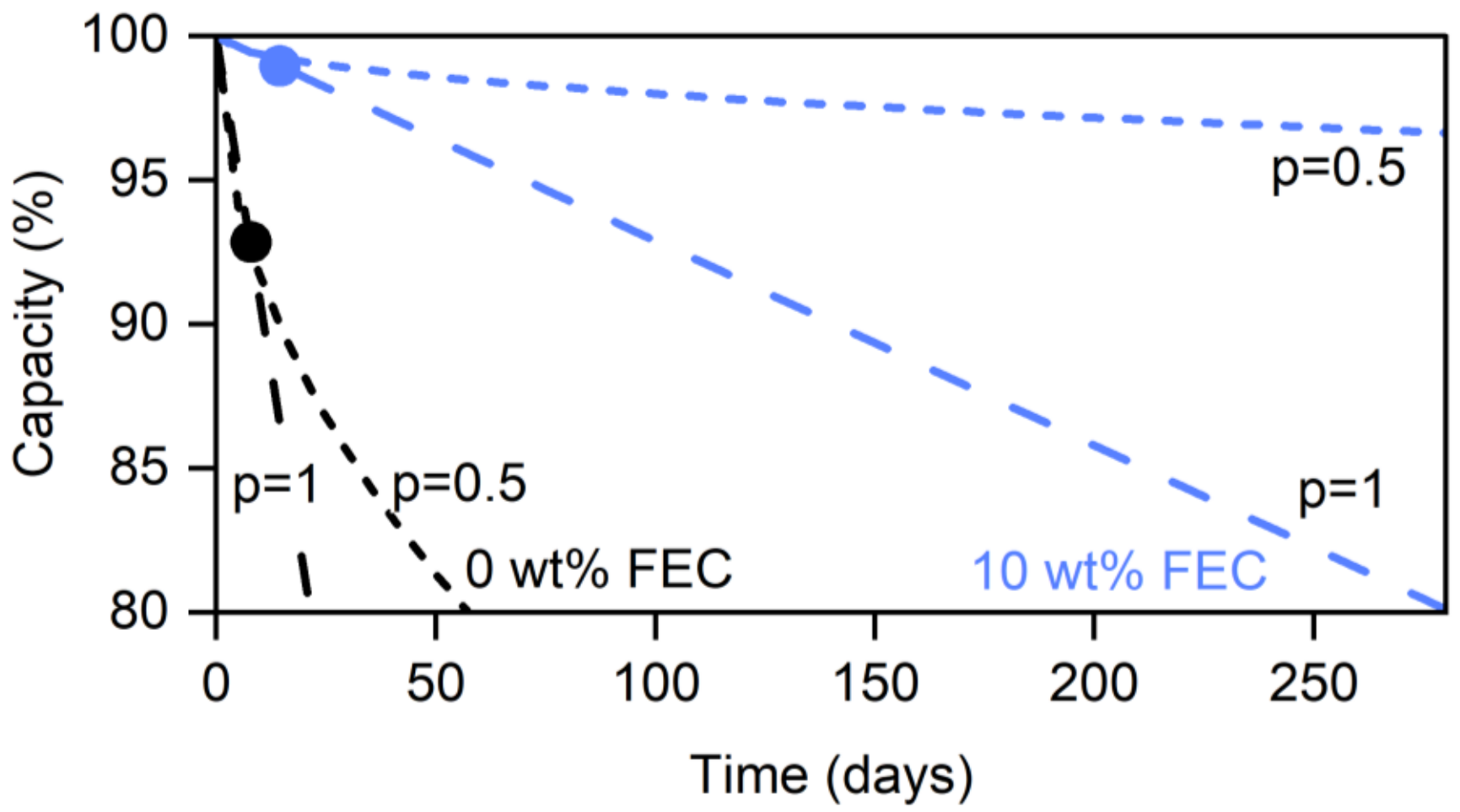

**Figure S2.** Prediction of calendar life based on the measured irreversible capacity loss during storage (solid data points). The curves show how capacity fades over time assuming two different degradation behaviors: p = 0.5, representing slower, diffusion-controlled aging, and p = 1, representing faster, reaction-controlled aging.

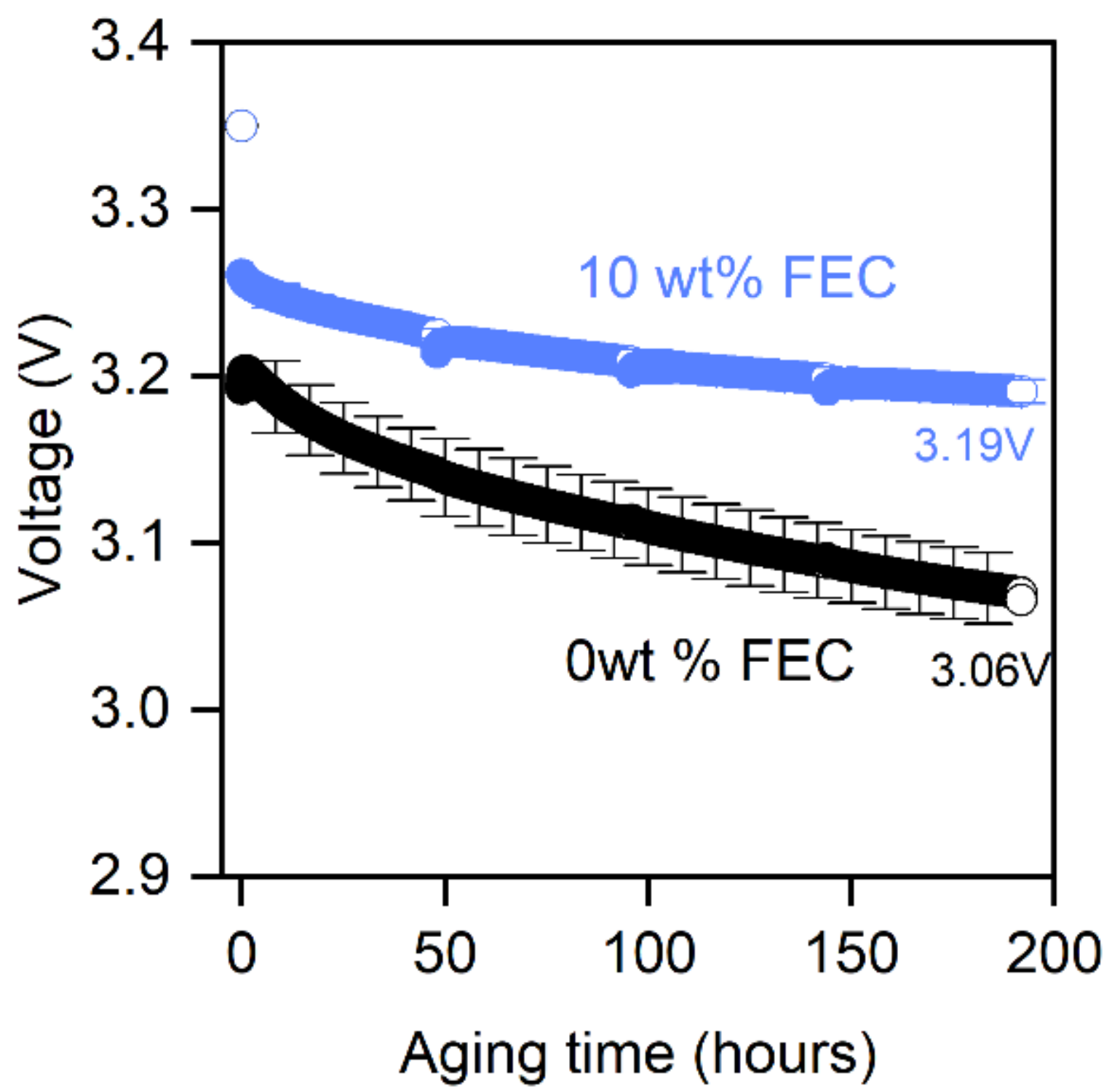

**Figure S3.** Open circuit voltage of 0wt% FEC and 10wt% FEC cell during 192 hours ( 8 days) aging period.

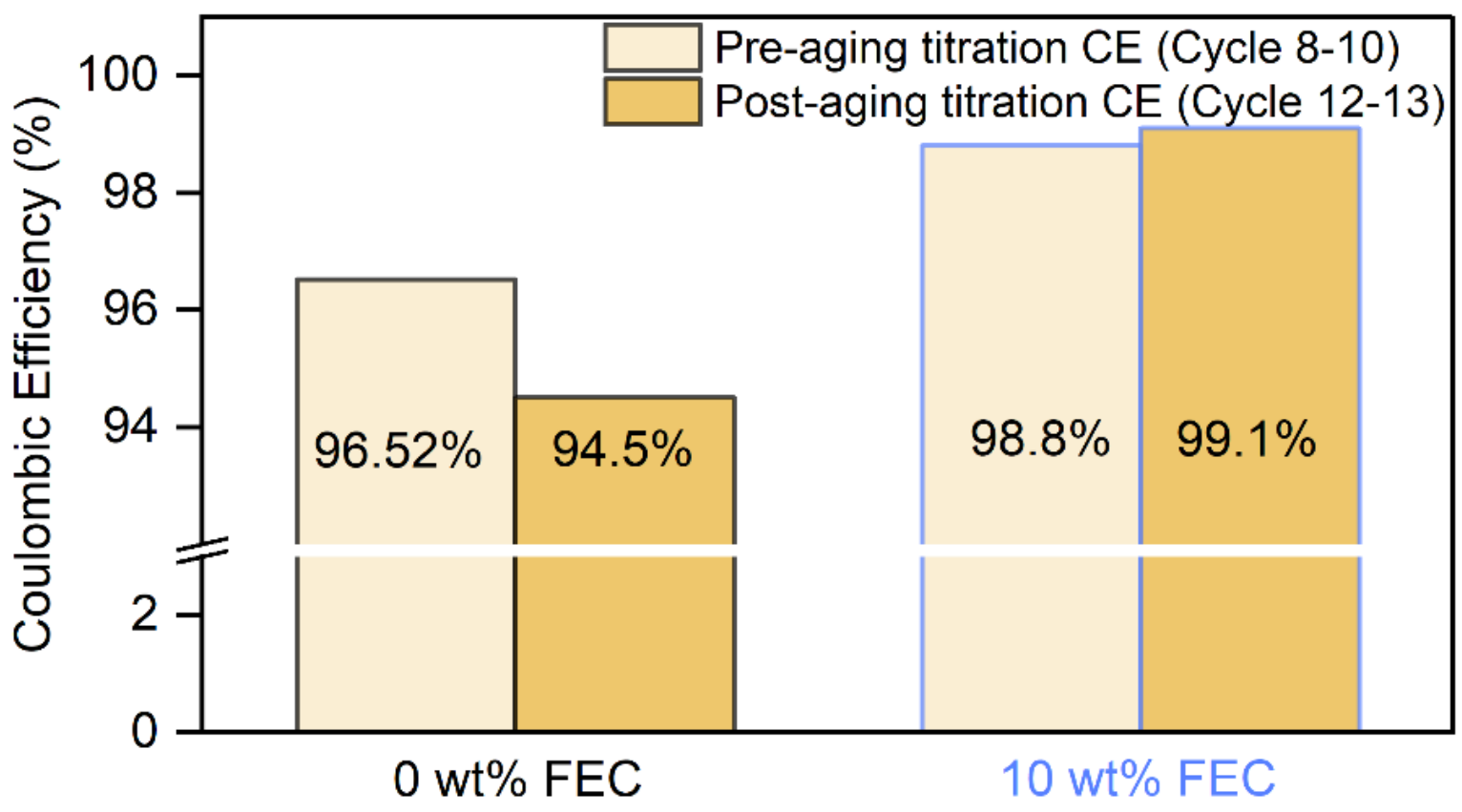

**Figure S4.** Average Coulombic efficiency (CE) obtained from Pre- and Post-aging titration steps for 0 wt% FEC and 10 wt% FEC cells after 192 hours (8 days) of aging.

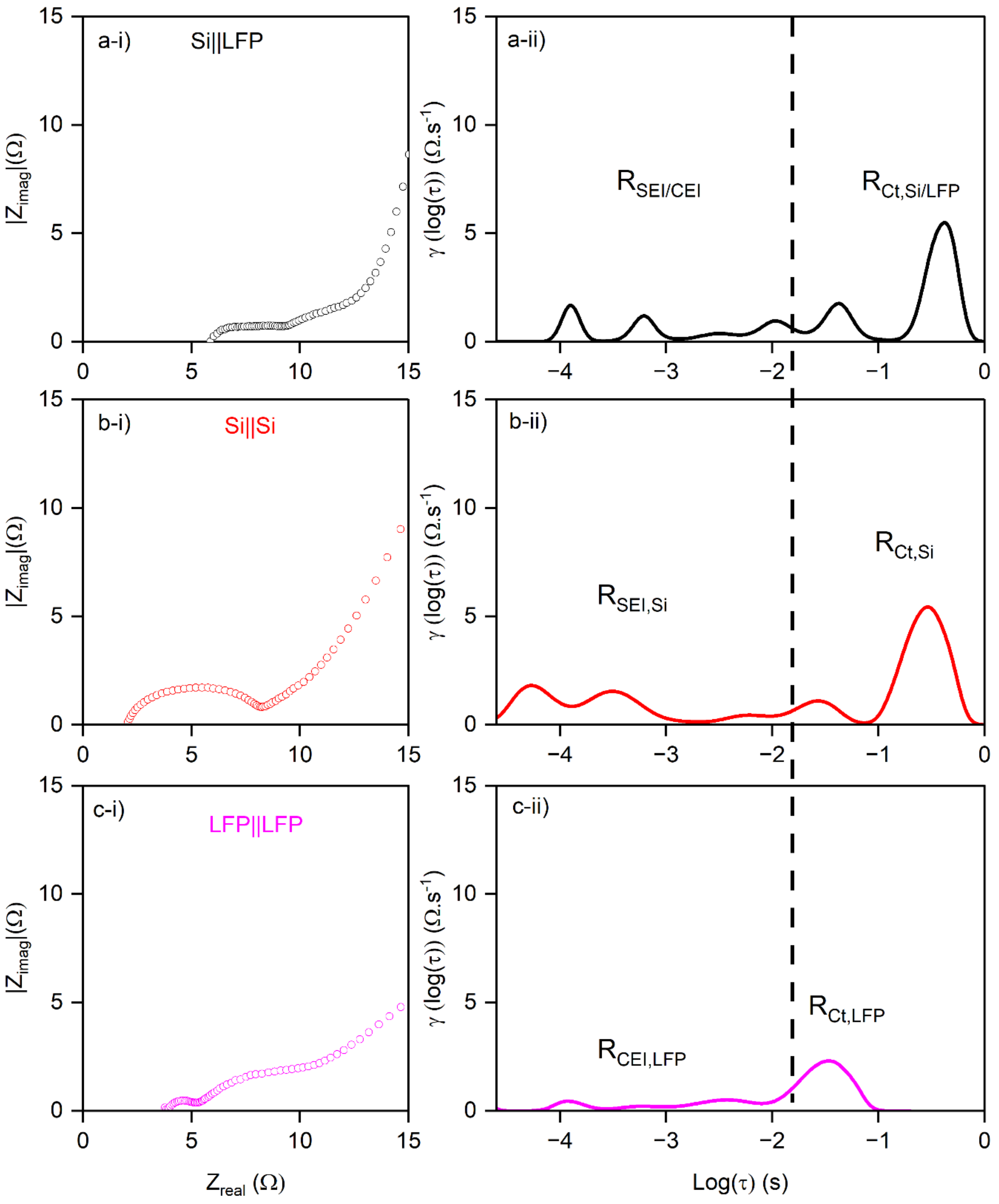

**Figure S5.** DRT peak identification of Si||LFP cells with Gen2 electrolyte using symmetric cells. (a–c)-i Nyquist plots from EIS and (a–c)-ii corresponding DRT spectra for (a) Si||LFP, (b) Si||Si, and (c) LFP||LFP cells. For symmetric cells, both axis in Nyquist plot and y axis in DRT plots are halved to show the contribution of a single electrode.

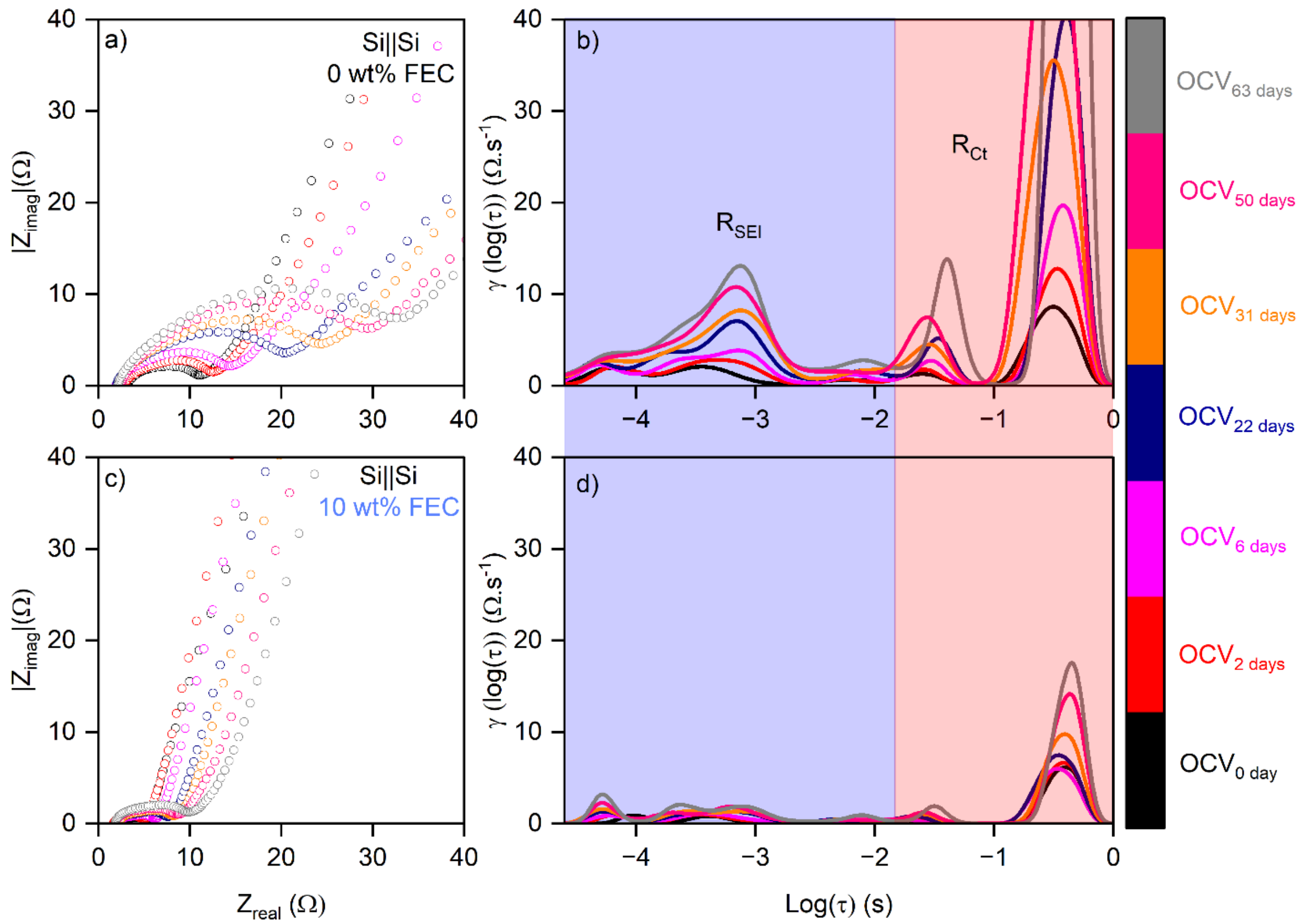

**Figure S6**. (a,c) Nyquist plots and (b,d) corresponding DRT spectra of Si||Si symmetric cells aged in Gen2 electrolyte without FEC (a,b) and with 10 wt% FEC (c,d).

**Table S2**. Power-law fitting parameters for SEI and charge-transfer resistances of symmetric cells during 63 days of aging

| | $R_{SEI} = b \cdot t^m$ | | $R_{ct} = c \cdot t^n$ | |
|---|---|---|---|---|
| | 0-8 days | 22-63 days | 0-8 days | 22-63 days |
| 0 wt% FEC | b = 2.52, m= 0.33 ($R^2$ =0.98) | b = 1.00, m= 0.62 ($R^2$ =0.99) | c = 4.17, n= 0.29 ($R^2$ =0.98) | c = 2.62, n= 0.52 ($R^2$ =0.99) |
| 10 wt% FEC | b = 0.96, m= 0.28 ($R^2$ =0.83) | b = 0.47, m= 0.49 ($R^2$ =0.96) | c = 3.59, n= 0.1 ($R^2$ =0.71) | c = 1.49, n= 0.38 ($R^2$ =0.98) |

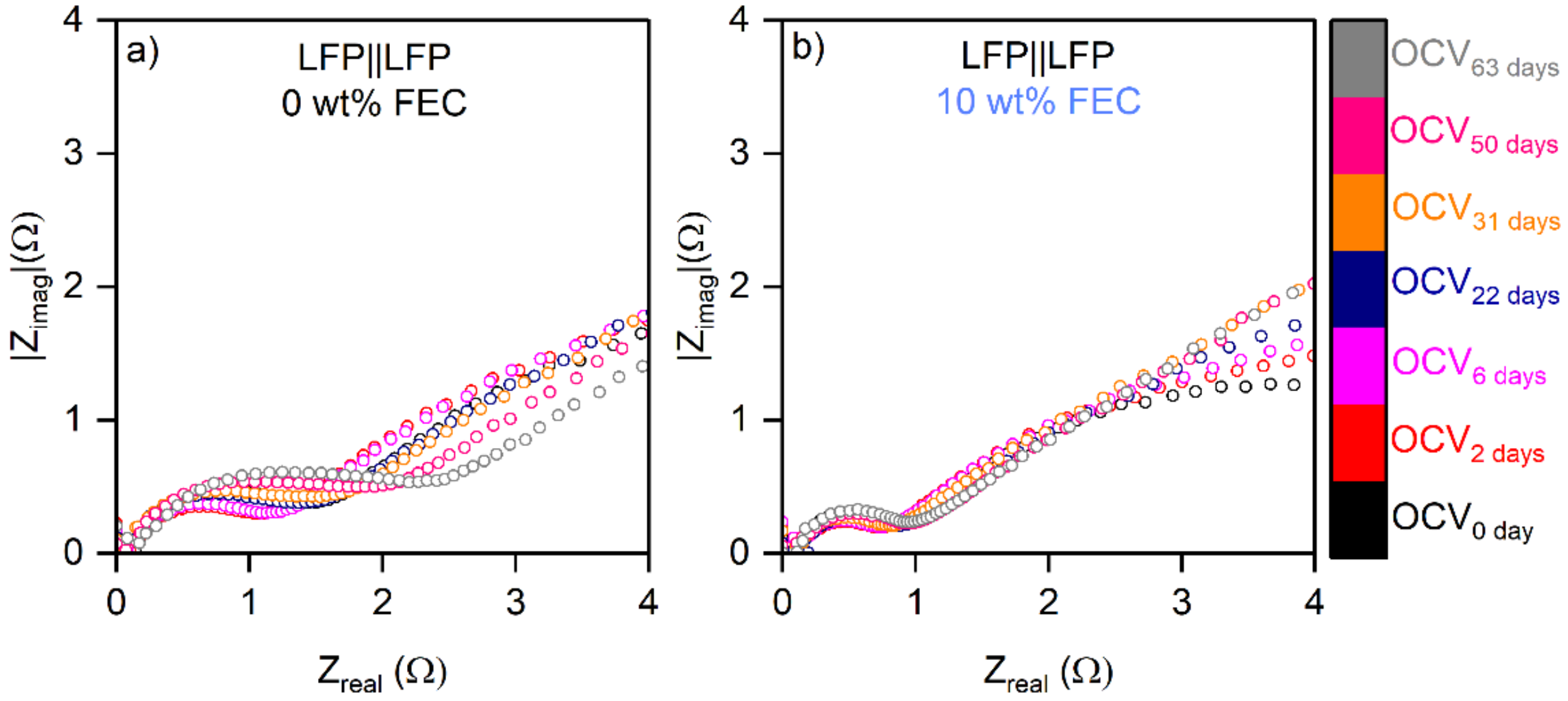

**Figure S7**. Nyquist plot obtained from EIS for LFP||LFP symmetric cell in (a) 0 wt% FEC and (b) 10wt% FEC condition.

**Table S2**: Summary of the XPS peaks of carbon ($C_{1s}$), oxygen ($O_{1s}$), lithium ($Li_{1s}$), fluorine ($F_{1s}$), and silicon ($Si_{2p}$) spectra.

| Spectra | Binding energy (eV) | Bonds | Reference |
|---|---|---|---|
| $C_{1s}$ | 285 | C-C | [1–3] |
| | 286.8 – 287.1 | C-O | [1,4,5] |
| | 288.9 – 289.8 | O-C=O | [5–7] |
| | 291.2 | C-F | [8] |
| $O_{1s}$ | 531.9 – 532.1 | C=O | [2,3,9] |
| | 533.2 – 533.5 | C-O, Si-O | [2,6] |
| | 534.6 | $CO_3$ (Organic) | [3] |
| $Li_{1s}$ | 53.5 – 53.7 | $Li_2O$ | [10] |
| | 55.2 – 55.4 | Li-O | [3,11,12] |
| | 56.1 – 56.3 | LiF | [13,14] |
| $F_{1s}$ | 685.1 – 685.2 | LiF | [2,5,9] |
| | 687.1 – 688.1 | $Li_xPO_yF_z$ | [2,5,15] |
| $Si_{2p}$ | 98 | $Li_xSi$ | [16,17] |
| | 102.2 – 102.4 | $Li_xSiO_y$ | [2,18] |
| | 103.8 | $Li_{x-p}SiO_q$ | [19–21] |

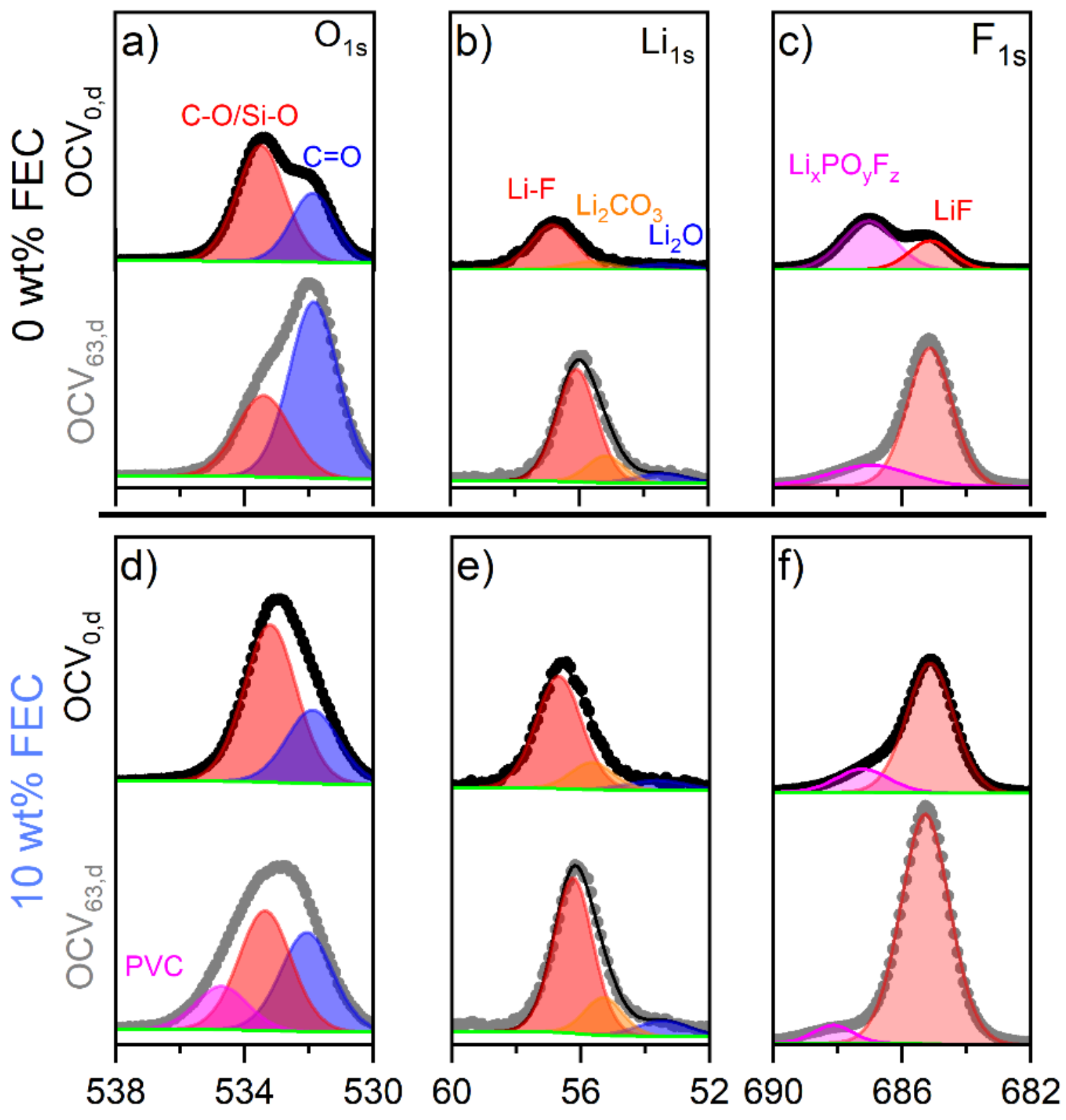

**Figure S8**. O1s, Li1s and F1s XPS for (a-c) 0 wt% FEC (d-f) 10wt% FEC cycled Si electrode before and after open circuit voltage (OCV) of 63 days.

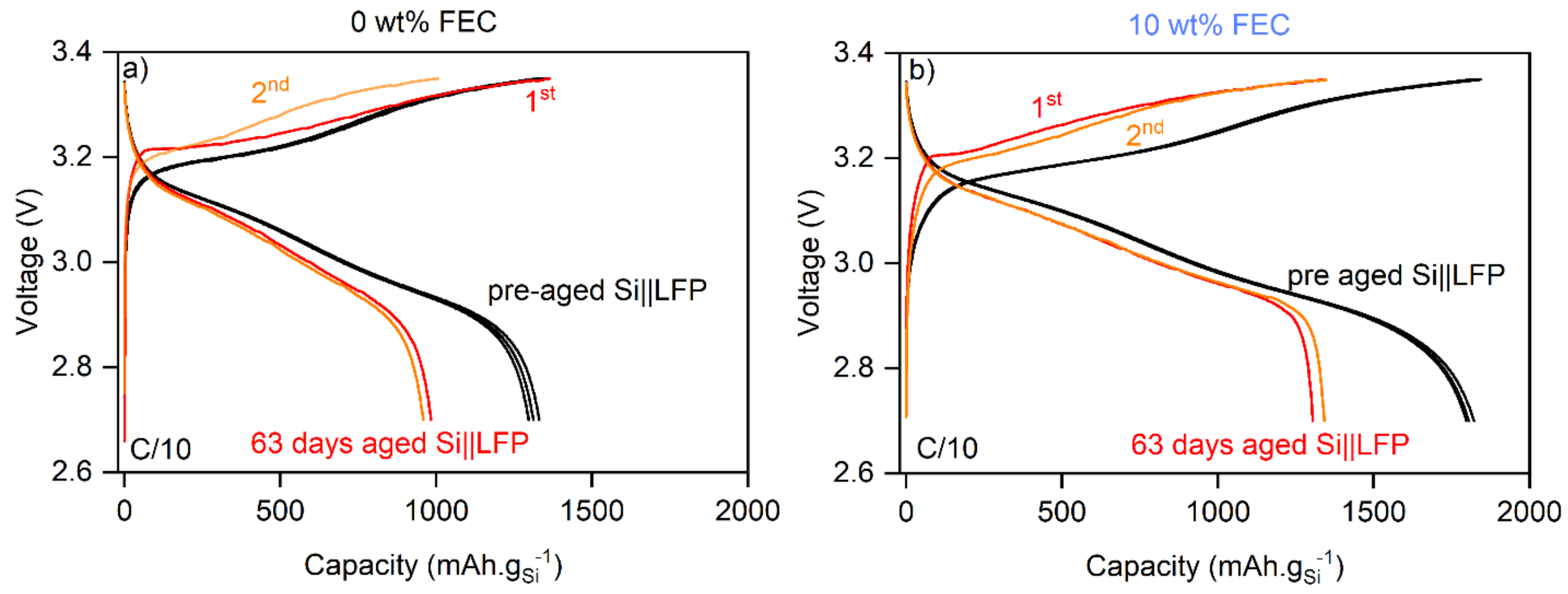

**Figure S9.** Voltage vs. Capacity plot of (a) 0 wt% FEC and (b) 10 wt% FEC Si||LFP full cells acquired during C/10 cycling. For each electrolyte, voltage profiles are shown for pre-aged Si

electrodes (cycles 8–10 from the pre-aging titration step in Figure 1) and for Si electrodes aged for 63 days in symmetric cell (first two cycles after reassembly).

SUPPORTING INFORMATION REFERENCES